\documentclass[fleqn,usenatbib]{mnras}

\usepackage[T1]{fontenc}
\usepackage{ae,aecompl}

\usepackage{graphicx}	% Including figure files
\usepackage{amsmath}	% Advanced maths commands
\usepackage{amssymb}	% Extra maths symbols
\usepackage[T1]{fontenc} 
\usepackage{academicons}
\usepackage{xcolor}
\usepackage{hyperref}

\usepackage{caption}
\usepackage{subcaption}

\usepackage{rotating}
\newcommand{\orcid}[1]{\href{https://orcid.org/#1}{\textcolor[HTML]{A6CE39}{\aiOrcid}}}

\def\ltsima{$\buildrel<\over\sim$}
\def\lsim{\lower.5ex\hbox{\ltsima}~}
\def\gtsima{$\buildrel>\over\sim$}
\def\gsim{\lower.5ex\hbox{\gtsima}~}

\def\teff{\ifmmode T_{\rm eff} \else $T_{\mathrm{eff}}$\fi}

\def\cm2{cm$^{-2}$}

\def\ewo3{$EW_{\mathrm{[O\textsc{iii}]}}$}

\def\nh{\ifmmode N_{\mathrm{HI}}\else $N_{\mathrm{HI}}$\fi}

\def\vexp{\ifmmode v_{\rm exp} \else v$_{\rm exp}$\fi}
\def\taua{\ifmmode \tau_{a}\else $\tau_{a}$\fi}

\usepackage{newtxtext,newtxmath}
\title[$z\sim9-16$ galaxy physical parameters]{Constraining the physical properties of the first lensed $z\sim9-16$ galaxy candidates with JWST}

\author[Furtak et al.]{Lukas J. Furtak$^{1}$\thanks{E-mail: furtak@post.bgu.ac.il},
Marko Shuntov$^{2}$,
Hakim Atek$^{2}$,
Adi Zitrin$^{1}$,
Johan Richard$^{3}$,
\newauthor Matthew D. Lehnert$^{3}$,
 and Jacopo Chevallard$^{4}$
\\
$^{1}$Physics Department, Ben-Gurion University of the Negev, P.O. Box 653, Be’er-Sheva 84105, Israel\\
$^{2}$Institut d'Astrophysique de Paris, CNRS, Sorbonne Universit\'e, 98bis Boulevard Arago, 75014, Paris, France\\
$^{3}$Univ Lyon, Univ Lyon1, Ens de Lyon, CNRS, Centre de Recherche Astrophysique de Lyon UMR5574, F-69230, Saint-Genis Laval, France\\
$^{4}$Department of Physics, University of Oxford, Denys Wilkinson Building, Keble Road, Oxford OX1\,3RH, UK\\
}

\date{Accepted 2022 December 14. Received 2022 December 14; in original form 2022 August 12}

\pubyear{2022}
\begin{document}
\label{firstpage}
\pagerange{\pageref{firstpage}--\pageref{lastpage}}
\maketitle

\begin{abstract}
The first deep-field observations of the JWST have immediately yielded a surprisingly large number of very high redshift candidates, pushing the frontier of observability well beyond $z\gtrsim10$. We here present a detailed SED-fitting analysis of the 10 gravitationally lensed $z\sim9-16$ galaxy candidates detected behind the galaxy cluster SMACS~J0723.3-7327 in a previous paper using the \texttt{BEAGLE} tool. Our analysis makes use of dynamical considerations to place limits on the ages of these galaxies and of all three published SL models of the cluster to account for lensing systematics. We find the majority of these galaxies to have relatively low stellar masses $M_{\star}\sim10^7-10^8\,\mathrm{M}_{\odot}$ and young ages $t_{\mathrm{age}}\sim10-100$\,Myr but with a few higher-mass exceptions ($M_{\star}\sim10^9-10^{10}\,\mathrm{M}_{\odot}$) due to Balmer-break detections at $z\sim9-10$. Because of their very blue UV-slopes, down to $\beta\sim-3$, all of the galaxies in our sample have extremely low dust attenuations $A_V\lesssim0.02$. Placing the measured parameters into relation, we find a very shallow $M_{\star}-M_{\mathrm{UV}}$-slope and high sSFRs above the main sequence of star-formation with no significant redshift-evolution in either relation. This is in agreement with the bright UV luminosities measured for these objects and indicates that we are naturally selecting UV-bright galaxies that are undergoing intense star-formation at the time they are observed. Finally, we discuss the robustness of our high-redshift galaxy sample regarding low-redshift interlopers and conclude that low-redshift solutions can safely be ruled out for roughly half of the sample, including the highest-redshift galaxies at $z\sim12-16$. These objects represent compelling targets for spectroscopic follow-up observations with JWST and ALMA.
\end{abstract}


%%%%%%%%%%%%%%%%%%%%%%%%%%%%%%%%%%%%%%%%%%%%%%%%%%

%%%%%%%%%%%%%%%%% BODY OF PAPER %%%%%%%%%%%%%%%%%%

\section{Introduction} \label{sec:intro}
The advent of the JWST has initiated a new era in high-redshift galaxy observations. Due to its unprecedented near-infrared (NIR) sensitivity, the JWST spectacularly expanded the frontier of observability beyond the $z\gtrsim10$ limit of the \textit{Hubble Space Telescope} (HST) and enables us to observe out to the first generation of galaxies -- the very first luminous structures that formed in the the Universe. The on-set of galaxy formation in the Universe is believed to have taken place at $z>15$. Observing the first stars and galaxies thus represents one of the fundamental challenges in modern astronomy and is one of the primary missions of the JWST.

Indeed, in the short time since the release of the first scientific observations of the JWST, the Early Release Observations \citep[ERO;][]{pontoppidan22} and the Early Release Science (ERS) programs GLASS-JWST \citep[PI: T. Treu;][]{treu22} and CEERS \citep[PI: S. Finkelstein;][]{finkelstein23,bagley22} have already yielded the first detections of galaxies beyond $z\gtrsim10$: Thus far, up to 11 bright galaxy candidates at $z\sim10-18$ were detected in the blank fields, GLASS-JWST and CEERS \citep[][]{naidu22b,castellano22,donnan23,finkelstein22,labbe22}, and up to 12 $z\sim9-16$ candidates \citep[][]{adams23,atek23} in the strong lensing (SL) cluster SMACS~J0723.3-7327 (SMACS0723) which had previously been imaged with the HST as part of the \textit{Reionization Lensing Cluster Survey} \citep[RELICS;][]{coe19}. There have even been some tentative detections out to $z\sim20$ \citep[][]{yan22}. These recent detections have also already yielded the first estimates of the rest-frame ultra-violet (UV) luminosity function \citep[][]{donnan23,harikane22}, galaxy physical parameters \citep[e.g.][]{labbe22,whitler23,topping22,rodighiero23,cullen22} and even possible implications for the star-formation history (SFH) of the Universe \citep[e.g.][]{mason22,boylan-kolchin22}. This demonstrates the formidable capability of the JWST to probe galaxies in the early Universe and foreshadows the numerous detections to come with the planned deep imaging programs of both SL clusters and blank fields. Caution is, however, also in order, since new populations of possible low-redshift interlopers in these kinds of studies have also already been identified \citep[][]{nonino22,fudamoto22,nelson22,barrufet22,zavala22,naidu22c,glazebrook22}. Moreover, while JWST has already delivered the first rest-frame optical spectroscopy of galaxies at $z\lesssim9$ which allows us to robustly characterize the stellar populations in these galaxies \citep[e.g.][]{carnall23,schaerer22b,laporte22,roberts-borsani22,williams22}, we remain limited to imaging data for galaxies $z\gtrsim10$. This means that we mostly observe these galaxies with rest-frame UV broad-band photometry which has been shown to be prone to parameter degeneracies and to not be well-suited for probing galaxy parameters in high-redshift studies with HST \citep[e.g.][]{grazian15,furtak21}. In order to infer the physical processes occurring within the first galaxies at $z\gtrsim10$ with JWST photometry, we therefore need to implement the lessons learned from HST observations and carefully account for parameter degeneracies and uncertainties.

In our previous work, \citet{atek23}, we presented the detection of 10 lensed $z\sim9-16$ galaxy candidates in SMACS0723 via the dropout selection technique in the JWST ERO observations of the cluster and measured their photometry, photometric redshifts and a first estimate of some galaxy parameters such as e.g., stellar mass. In order to address the issues explained above and derive first robust parameter estimations with JWST, we present here an in-depth analysis of the spectral energy distributions (SEDs) of this high-redshift galaxy sample with the Bayesian tool \texttt{BEAGLE} \citep{chevallard16}. Using all the information currently available (e.g., photometric and morphological measurements, lensing models, etc.), we make a first assessment of what can be learned of these objects with JWST and what are the uncertainties involved. The goal of this study is also to establish a method to estimate physical parameters at high redshifts which will no doubt be of use to future JWST surveys set to observe these kinds of galaxies in the early Universe.

This paper is organized as follows: We present our sample and derivation of physical parameters in section~\ref{sec:SED-fit}. We then place our resulting galaxy parameters in relation to each other in section~\ref{sec:relations} and discuss our results in section~\ref{sec:discussion}. Finally, we summarize our analysis and findings in the conclusions, in section~\ref{sec:conclusion}. Throughout this paper, we assume a standard flat $\Lambda$CDM cosmology with $H_0=70\,\frac{\mathrm{km}}{\mathrm{s}\,\mathrm{Mpc}}$, $\Omega_{\Lambda}=0.7$, and $\Omega_\mathrm{m}=0.3$. All magnitudes quoted are in the AB system \citep{oke83} and all quoted uncertainties represent $1\sigma$ ranges.

\section{Galaxy parameters} \label{sec:SED-fit}
In this study, we use the $z\sim9-16$ galaxy sample detected in \citet{atek23} which is composed of 6 candidates in the redshift range $z\sim9-11$ and 4 candidates at $z\gtrsim12$, up to $z\sim16$. For this sample, observations in seven broad-band filters are available: the F090W, F150W, F200W, F277W, F356W and F444W bands from the \textit{Near-Infrared Camera} \citep[NIRCam;][]{rieke05} and the F115W band from the \textit{Near-Infrared Imager and Slitless Spectrograph} \citep[NIRISS;][]{doyon12} in a smaller field-of-view centered on the cluster core. The photometry was measured with \texttt{SExtractor++} \citep[\texttt{SE++};][]{bertin20,kuemmel20}. We refer the reader to \citet{atek23} for the details of data reduction, source detection and photometry with \texttt{SE++}, and the high-redshift dropout selection methods used for our sample.

In order to overcome some of the degeneracies inherent to fitting multiple galaxy parameters to only a hand-full of broad-band rest-frame UV photometric filters, we use all of the information available to place as many constraints on the galaxy parameters as possible. This is done in section~\ref{sec:priors}, before performing an SED-fit in section~\ref{sec:BEAGLE-setup} and correcting the resulting stellar mass for gravitational magnification in section~\ref{sec:SL}.

\subsection{External priors} \label{sec:priors}
In \citet{atek23}, we measured UV-continuum slopes $\beta$ and half-light-radii $r_e$ for each object (cf. Tab.~2 in \citealt{atek23}), which will both come in handy to constrain the physical parameters of our high-redshift galaxies.

We first use the well known relation between the UV-continuum slope and the dust reddening $E(B-V)$ \citep[e.g.][]{meurer99,reddy18a} to compute the effective \textit{V}-band dust attenuation optical depth needed in our SED-fit (cf. section~\ref{sec:BEAGLE-setup}) for each galaxy in our sample. For that we adopt the relation measured by \citet{reddy18a} for an SMC-like dust attenuation curve:

\begin{equation} \label{eq:ebv}
    E(B-V)=\frac{\beta-\beta_0}{11.259}
\end{equation}

where $\beta_0=-2.616$ is the intrinsic slope measured in \citet{reddy18a}. We then use $R_V=2.505$ \citep[][]{reddy15} to convert the reddening to optical depth $\hat{\tau}_V$. We obtain very low values $\hat{\tau}_V\lesssim0.01$ due to the very blue UV-slopes of our sample \citep[cf.][]{atek23} and the relatively steep slope of the SMC law. Using slightly different intrinsic slopes \citep[e.g. the $\beta_0=-2.23$ from][]{meurer99} and values of $R_V$, we therefore do not find significantly different $\hat{\tau}_V$. Note that this is in agreement with the expectation for high-redshift galaxies to have very low dust attenuation and rules out the extreme dust attenuation scenarios discussed in \citet{rodighiero23}.

Next, we use the (lensing-corrected) half-light-radii and first stellar mass estimates (cf. Tab~2 in \citealt{atek23}) to compute the dynamical time scale of each galaxy \citep[see also, e.g., discussion in][]{verma07} as,

\begin{equation} \label{eq:tdyn}
    t_{\mathrm{dyn}}\sim\frac{r}{v}\sim\sqrt{\frac{2r^3}{GM_{\star}}}
\end{equation}

\noindent This does not necessarily need to be a precise measurement but rather a rough estimate of order of magnitude. Following our approach in \citet{furtak21}, we will use this estimate to place a lower boundary on the range of allowed stellar ages in the SED-fit, thus assuming that a galaxy cannot be younger than the lower range of these dynamical timescales, and to place a prior on the characteristic star-formation timescale. We overall find our sample of high-redshift galaxies to have small dynamical times of the order $\sim2-10$\,Myr which is not surprising given their very compact morphologies \citep[$r_e\lesssim0.6$\,kpc for the majority, cf.][]{atek23}.

Note that throughout this analysis, we propagate the uncertainties of $\beta$ and $r_e$ to the parameters computed in~\eqref{eq:ebv} and~\eqref{eq:tdyn} so that they can be properly accounted for in the SED-fit.

\subsection{SED-fit setup} \label{sec:BEAGLE-setup}
Said SED-fit is performed with the \texttt{BayEsian Analysis of GaLaxy sEds} \citep[\texttt{BEAGLE};][]{chevallard16} tool, the fully Bayesian nature of which is ideally suited to optimize numerous galaxy parameters at once, including priors, and robustly probe and combine their uncertainties in the joint posterior probability distribution function (PDF). It uses SED templates by \citet{gutkin16} which combine the latest version of the stellar population synthesis models by \citet{bc03} with the photoionization code \texttt{CLOUDY} \citep[][]{ferland13} to account for nebular emission. These templates all assume a \citet{chabrier03} initial mass function (IMF) and the latest models of intergalactic medium (IGM) attenuation by \citet{inoue14}.

For the fit, we assume a delayed exponential star-formation history (SFH) $\psi\propto t\exp(-t/\tau)$ with the possibility of an ongoing star-burst over the last 10\,Myr. This allows for maximum flexibility of the SFH to be either rising or declining with a maximum at $t=\tau$. We furthermore assume an SMC-like dust attenuation law \citep[][]{pei92} which has been found to match high-redshift galaxies best \citep[][]{capak15,reddy15,reddy18a}, in particular in the low-metallicity regime \citep[][]{shivaei20}.

With this setup, we fit four free parameters:

\begin{itemize}
    \item Stellar mass $M_{\star}$ with a log-uniform prior $\log(M_{\star}/\mathrm{M}_{\odot})\in[6,10]$.
    \item Star-formation rate (SFR) $\psi$ over the last $10^7$\,yr with a log-uniform prior $\log(\psi/\mathrm{M}_{\odot}\,\mathrm{yr}^{-1})\in[-4,4]$.
    \item Maximum stellar age $t_{\mathrm{age}}$ with a log-uniform prior $\log(t_{\mathrm{age}/\mathrm{yr}})\in[6.3, t_{\mathrm{universe}}]$ where the lower boundary was chosen according to our analysis of the dynamical time (cf. section~\ref{sec:priors}) and the upper boundary is the age of the Universe at the redshift of the galaxy.
    \item Stellar metallicity $Z$ with a log-uniform prior $\log(Z/\mathrm{Z}_{\odot})\in[-2,-0.3]$.
\end{itemize}

For the other parameters necessary for the fit we use the values measured independently by setting Gaussian priors with the measured values as the mean and the measured uncertainties as the standard deviation:

\begin{itemize}
    \item Characteristic star-forming time scale $\tau$ with the dynamical time $t_{\mathrm{dyn}}$ computed in section~\ref{sec:priors} as prior.
    \item Effective \textit{V}-band dust attenuation optical depth $\hat{\tau}_V$ with the attenuation inferred from the UV-slope with Eq.~\eqref{eq:ebv} as prior.
    \item Photometric redshift $z_{\mathrm{phot}}$ with the value measured in \citet{atek23} as prior.
\end{itemize}

This method allows for the uncertainties on the measured parameters to propagate to the SED-fit and into the posterior PDF. We run the SED-fit on all seven bands of broad-band photometry available for our galaxy sample.

\subsection{Gravitational magnification} \label{sec:SL}

\begin{table}
    \centering
    \caption{Gravitational magnifications of our high-redshift objects in the three JWST-based SL models of SMACS0723: \citet{mahler22} (M22), \citet{pascale22} (P22) and \citet{caminha22} (C22).}
    \begin{tabular}{lcccc}
        \hline
        ID          &   $\mu_{\mathrm{M22}}$    &   $\mu_{\mathrm{P22}}$    &    $\mu_{\mathrm{C22}}^{\mathrm{a}}$  &   $\mu^{\mathrm{b}}$\\\hline
        SMACS\_z10a &   $3.98\pm0.30$  &   $6.14\pm1.74$           &    $8.78$                             &   $6.30\pm2.21$\\
        SMACS\_z10b &   $1.36\pm0.03$  &   $1.70\pm0.14$           &    $1.70$                             &   $1.56\pm0.18$\\
        SMACS\_z10c &   $1.41\pm0.03$  &   $1.70\pm0.14$           &    $1.69$                             &   $1.60\pm0.16$\\
        SMACS\_z10d &   $1.13\pm0.01$  &   $1.35\pm0.07$           &    $1.34$                             &   $1.28\pm0.11$\\
        SMACS\_z10e &   $1.07\pm0.01$  &   $1.24\pm0.04$           &    $1.22$                             &   $1.18\pm0.08$\\
        SMACS\_z11a &   $1.05\pm0.01$  &   $1.22\pm0.04$           &    $1.20$                             &   $1.16\pm0.08$\\
        SMACS\_z12a &   $1.05\pm0.01$  &   $1.21\pm0.04$           &    $1.18$                             &   $1.15\pm0.07$\\
        SMACS\_z12b &   $1.35\pm0.03$  &   $1.65\pm0.13$           &    $1.64$                             &   $1.55\pm0.16$\\
        SMACS\_z16a &   $1.86\pm0.05$  &   $2.41\pm0.29$           &    $2.27$                             &   $2.18\pm0.29$\\
        SMACS\_z16b &   $1.04\pm0.01$  &   $1.19\pm0.03$           &    $1.17$                             &   $1.13\pm0.07$\\\hline
    \end{tabular}
    \par\smallskip
    \begin{flushleft} $^{\mathrm{a}}$\, Uncertainties for the \citet{caminha22} model are not publicly available.
    \par $^{\mathrm{b}}$\, Average magnification used for parameter estimation in this study (cf. section~\ref{sec:SL} for details).
    \end{flushleft}
    \label{tab:SL}
\end{table}

Since we are observing sources behind a lensing cluster, SMACS0723, we need to account for the gravitational magnification and correct certain parameter measurements measurements (cf. section~\ref{sec:SED-fit_results}). Other parameters estimated in sections~\ref{sec:priors} and~\ref{sec:BEAGLE-setup} depend on relative fluxes, i.e. colors, and are therefore not affected by the achromatic SL magnification.

The JWST ERO of SMACS0723 yielded numerous new multiple image systems to constrain the SL models of the cluster, increasing the number of 5 multiple image systems known from HST observations \citep[][]{golubchik22} up to more than 20. To date, there are three SL models of SMACS0723 based on the new JWST observations: A parametric model built with \texttt{lenstool} \citep[][]{kneib96,jullo07,jullo09} by \citet{mahler22}, an analytic model built with a revised version of the \citet{zitrin15a} parametric implementation, by \citet{pascale22}, and another \texttt{lenstool}-based parametric model by \citet{caminha22}.

We show the magnifications of our objects in each of the three JWST-based models in Tab.~\ref{tab:SL}. While the \citet{mahler22} model shows slightly lower magnifications than the \citet{pascale22} and \citet{caminha22} models, the overall scatter between models is relatively low. This is not surprising since none of our sources lies particularly close to a critical line where the impact of the modeling systematics is the most severe. Indeed, we do not have any extremely magnified objects in our sample which means that we are spared the worst of the lensing systematics that have been found to dominate the uncertainties in studies of lensed high-redshift galaxies with HST \citep[e.g.][]{bouwens17,bouwens22b,bouwens22a,atek18,furtak21}. In order to take into account all of the lensing information in this study, we follow a similar approach to the one of \citet{bhatawdekar19}, \citet{furtak21} and \citet{bouwens22b} and use the mean magnification $\mu$ from the three JWST-based SL models in the following. The magnification uncertainties $\Delta\mu$ are then computed by propagating the individual magnification uncertainties of each model, if available, and adding the scatter between the models in quadrature. The resulting values are also reported in Tab.~\ref{tab:SL}. With this approach we make sure that all sources of magnification uncertainties are accounted for and propagated to our final results.

\subsection{SED-fit results} \label{sec:SED-fit_results}

\begin{table*}
    \caption{\texttt{BEAGLE} SED-fit results for our whole $z\sim9-16$ sample. Best-fit results are taken as the median and the $1\sigma$-range of the joint posterior distribution of each galaxy. The last column contains the UV luminosities computed in section~\ref{sec:mass-light}.}
    \begin{tabular}{lcccccccc}
        \hline
        ID & $\log(M_{\star}/\mathrm{M}_{\odot})$ & $\log(\psi/\mathrm{M}_{\odot}\,\mathrm{yr}^{-1})$ & $\log(t_{\mathrm{age}/\mathrm{yr}})$ & $\log(Z/\mathrm{Z}_{\odot})$ & $\log(\tau/\mathrm{yr})^{\mathrm{a}}$ & $\hat{\tau}_V^{\mathrm{a}}$ & $z_{\mathrm{phot}}^{\mathrm{a}}$  &   $M_{\mathrm{UV}}^{\mathrm{b}}$\\\hline
        SMACS\_z10a &   $8.86_{-0.15}^{+0.16}$  &   $-1.97_{-0.54}^{+0.63}$ &   $7.96_{-0.04}^{+0.05}$  &   $-0.39_{-0.23}^{+0.07}$ &   $6.28\pm0.03$  &   $0.032\pm0.002$   &   $9.77_{-0.02}^{+0.02}$  &   $-18.3\pm0.4$\\
        SMACS\_z10b &   $10.21_{-0.05}^{+0.05}$ &   $-1.42_{-0.54}^{+0.61}$ &   $8.32_{-0.02}^{+0.02}$  &   $-1.19_{-0.02}^{+0.01}$ &   $6.44\pm0.02$  &   $0.037\pm0.002$   &   $9.03_{-0.01}^{+0.01}$  &   $-20.6\pm0.2$\\
        SMACS\_z10c &   $9.72_{-0.05}^{+0.05}$  &   $-0.33_{-1.06}^{+0.27}$ &   $8.27_{-0.04}^{+0.02}$  &   $-1.83_{-0.08}^{+0.13}$ &   $6.48\pm0.01$  &   $0.019\pm0.004$   &   $9.78_{-0.01}^{+0.01}$  &   $-20.1\pm0.2$\\
        SMACS\_z10d &   $6.95_{-1.40}^{+2.45}$  &   $0.54_{-0.05}^{+0.47}$  &   $6.96_{-0.54}^{+1.27}$  &   $-1.21_{-0.12}^{+0.11}$ &   $7.56\pm0.07$  &   $0.017\pm0.006$   &   $9.32_{-0.07}^{+0.07}$  &   $-19.6\pm0.2$\\
        SMACS\_z10e &   $6.87_{-1.35}^{+1.42}$  &   $1.16_{-0.04}^{+0.05}$  &   $7.70_{-0.69}^{+0.64}$  &   $-1.42_{-0.21}^{+0.17}$ &   $6.85\pm0.11$  &   $0.026\pm0.010$   &   $10.88_{-0.10}^{+0.11}$ &   $-18.8\pm0.3$\\
        SMACS\_z11a &   $6.46_{-1.06}^{+1.16}$  &   $0.77_{-0.03}^{+0.03}$  &   $7.78_{-0.61}^{+0.58}$  &   $-0.88_{-0.07}^{+0.09}$ &   $6.82\pm0.11$  &   $0.021\pm0.011$   &   $11.08_{-0.05}^{+0.06}$ &   $-18.4\pm0.4$\\
        SMACS\_z12a &   $8.27_{-0.10}^{+0.12}$  &   $-1.33_{-0.52}^{+0.63}$ &   $7.32_{-0.17}^{+0.18}$  &   $-0.57_{-0.32}^{+0.20}$ &   $7.26\pm0.10$  &   $0.007\pm0.004$   &   $12.16_{-0.08}^{+0.08}$ &   $-19.7\pm0.2$\\
        SMACS\_z12b &   $8.26_{-0.29}^{+0.29}$  &   $-0.98_{-0.79}^{+0.91}$ &   $7.75_{-0.49}^{+0.41}$  &   $-1.63_{-0.29}^{+0.38}$ &   $8.32\pm0.11$  &   $0.025\pm0.008$   &   $12.27_{-0.10}^{+0.09}$ &   $-19.9\pm0.2$\\
        SMACS\_z16a &   $8.02_{-2.28}^{+1.07}$  &   $1.22_{-1.96}^{+0.07}$  &   $7.45_{-0.49}^{+0.50}$  &   $-1.13_{-0.29}^{+0.38}$ &   $7.42\pm0.17$  &   $0.006\pm0.003$   &   $15.93_{-0.11}^{+0.11}$ &   $-20.4\pm0.2$\\
        SMACS\_z16b &   $7.89_{-1.99}^{+2.37}$  &   $1.76_{-0.31}^{+0.22}$  &   $6.54_{-0.18}^{+0.83}$  &   $-1.41_{-0.40}^{+0.36}$ &   $8.18\pm0.26$  &   $0.016\pm0.009$   &   $15.25_{-0.09}^{+0.09}$ &   $-20.9\pm0.2$\\\hline
    \end{tabular}
    \par\smallskip
    \begin{flushleft}
        $^{\mathrm{a}}$\, Parameter fit with a Gaussian prior corresponding to values (and their uncertainties) previously derived from independent analyses as detailed in sections~\ref{sec:priors} and~\ref{sec:BEAGLE-setup}.
        \par $^{\mathrm{b}}$\, Updated luminosities computed with the magnifications $\mu$ reported in Tab.~\ref{tab:SL}.
    \end{flushleft}
    \label{tab:galay_parameters}
\end{table*}

\begin{figure}
    \centering
    \includegraphics[width=\columnwidth, keepaspectratio=true]{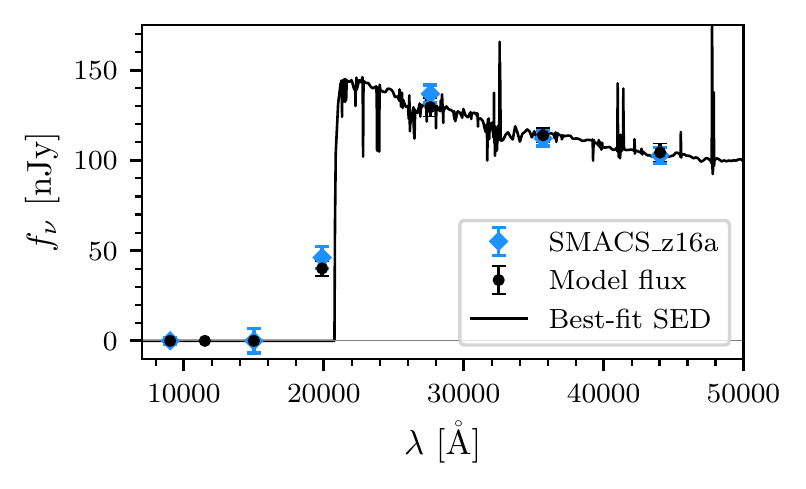}
    \includegraphics[width=\columnwidth, keepaspectratio=true]{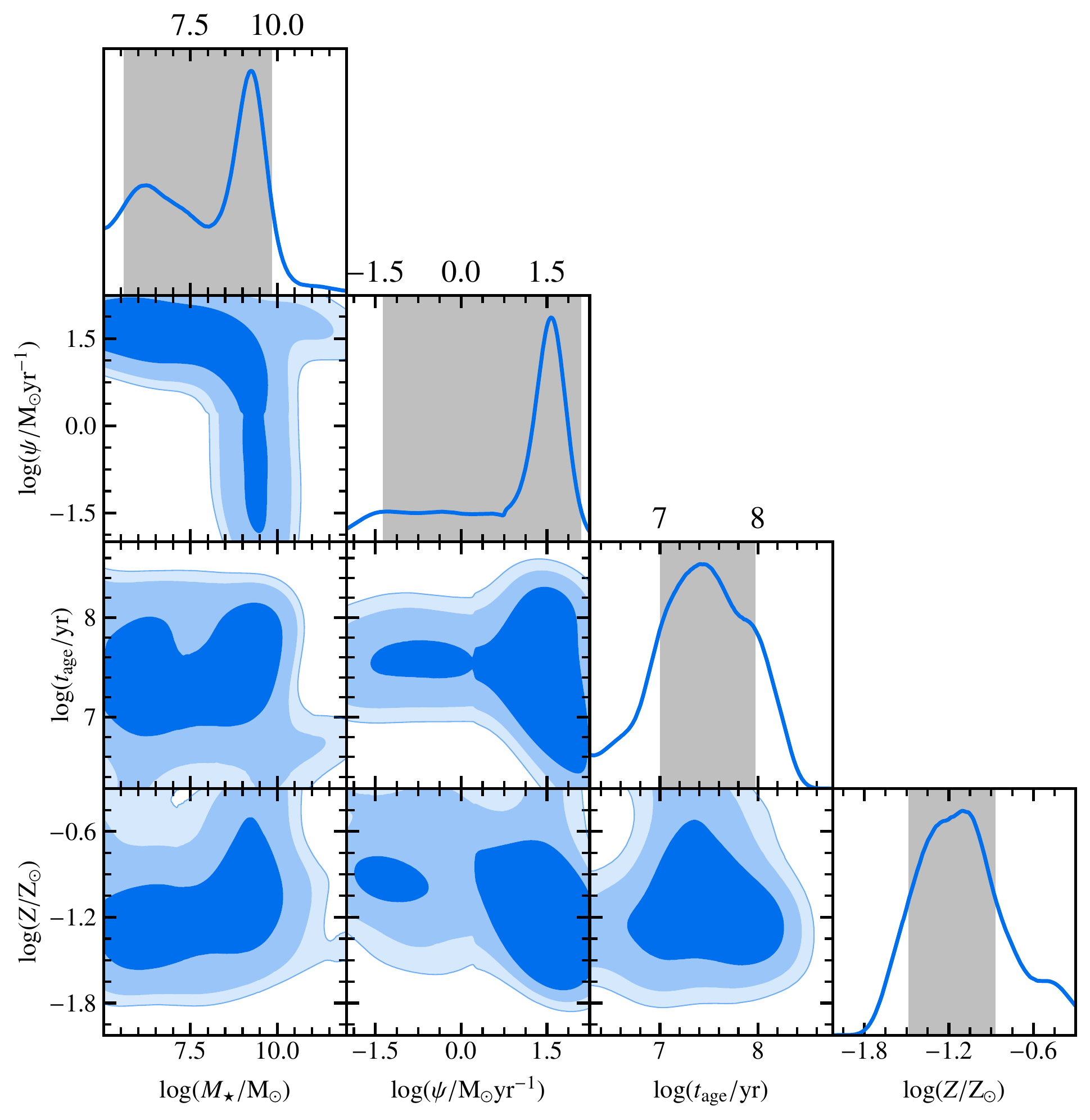}
    \caption{SED-fit results from \texttt{BEAGLE} for our highest redshift object, SMACS\_z16a at $z_{\mathrm{phot}}\simeq15.93\pm0.11$. \textit{Top panel}: Best-fitting (maximum-a-posteriori) SED (black) over the observed photometry (blue diamonds). \textit{Bottom panel}: Posterior distribution function for the four free parameters in our SED-fit: stellar mass $M_{\star}$, SFR $\psi$, maximum stellar age $t_{\mathrm{age}}$ and metallicity $Z$. Note that $M_{\star}$ and $\psi$ are not magnification corrected in this plot.}
    \label{fig:z16a_fit}
\end{figure}

\begin{figure}
    \centering
    \includegraphics[width=\columnwidth, keepaspectratio=true]{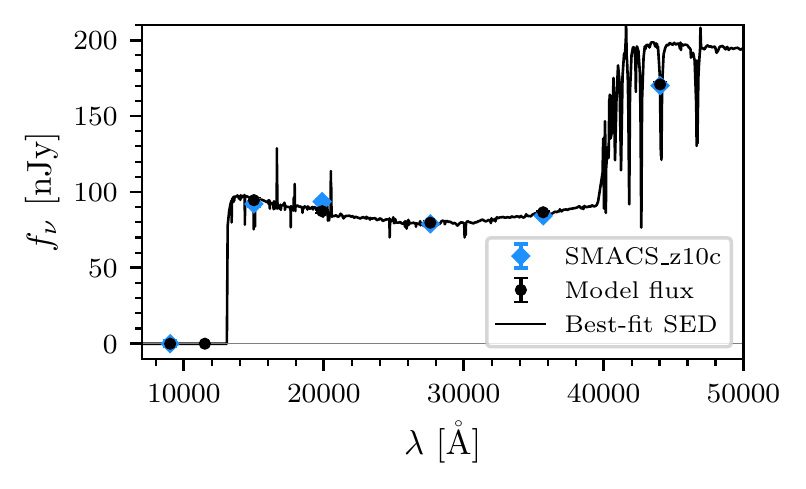}
    \includegraphics[width=\columnwidth, keepaspectratio=true]{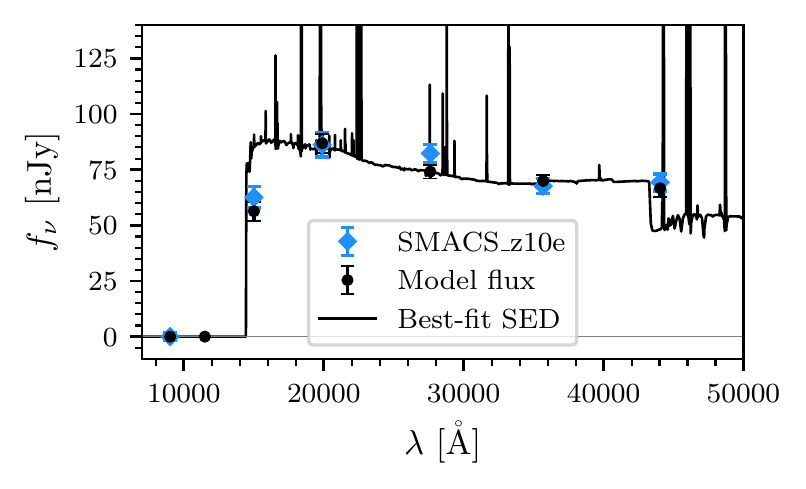}
    \caption{Same best-fit SED plot as in the top panel of Fig.~\ref{fig:z16a_fit} but for SMACS\_z10c (\textit{upper panel}) and for SMACS\_z10e (\textit{lower panel}). At $z\sim9-10$, the Balmer break is located in the F444W-band which allows to probe more evolved stellar populations and results in higher stellar masses and ages. Some galaxies at $z\sim10$ however do not show a pronounced Balmer-break which indicates them to have young ages and low stellar masses.}
    \label{fig:balmer_break}
\end{figure}

With the magnifications in hand, we correct the stellar masses and SFRs for the effects of lensing after the SED-fit. As was shown in \citet{furtak21}, this is equivalent to correcting the photometry prior to SED-fitting but has the advantage that we do not need to propagate the uncertainties of the magnification to the photometry but only to the parameters affected by it. 

The best-fit galaxy parameter values, taken as the median and $1\sigma$-range of the posterior distribution from our \texttt{BEAGLE} fit of each $z\sim9-16$ galaxy detected in SMACS0723, are shown in Tab.~\ref{tab:galay_parameters}. We also show an example of a best-fit SED and the posterior distribution of \texttt{BEAGLE} parameters for one object, SMACS\_z16a, in Fig.~\ref{fig:z16a_fit} as an example. As can be seen in Fig.~\ref{fig:z16a_fit} and in Tab.~\ref{tab:galay_parameters}, while the SED-fit is relatively good, there remain some parameter degeneracies between the stellar mass, the SFR and the age which result in the relatively large uncertainties on these parameters. This is due to the fact that we are fitting rest-frame UV photometry exclusively for the majority of our sample which is not ideally suited to probe galaxy parameters \citep[cf. e.g.][]{furtak21} as will be discussed in more detail in section~\ref{sec:BEAGLE_limits}. There is however a clearly defined maximum-a-posteriori solution for each parameter.

We overall find these objects to have relatively low stellar masses $M_{\star}\sim10^7-10^8\,\mathrm{M}_{\odot}$, down to $M_{\star}\simeq10^{6.5}\,\mathrm{M}_{\odot}$ which corresponds to the lowest stellar masses ever detected at moderate redshifts ($z\sim6-7$) with HST \citep[e.g.][]{bhatawdekar19,kikuchihara20,furtak21}. There are however also a few high-mass $M_{\star}\sim10^9-10^{10}\,\mathrm{M}_{\odot}$ objects on the low-redshift end of our sample. We also find relatively young ages $t_{\mathrm{age}}\sim10-100$\,Myr and high SFRs for most of our sample which is not surprising given the very blue UV-slopes of these galaxies \citep{atek23} and in agreement with other studies of the first observations of the JWST high-redshift frontier \citep[e.g.][]{nanayakkara22,whitler23,topping22}. The SFR-$M_{\star}$ relation and its implications will be discussed further in section~\ref{sec:mass-sfr}. While the metallicity is in general not very well constrained, there is a strong tendency towards $Z<0.1\,\mathrm{Z}_{\odot}$ in most galaxies of our sample. We will discuss the limits of our SED-fits with regards to stellar age and metallicity in detail in section~\ref{sec:BEAGLE_limits}.

We however also note the presence of several older objects with ages $t_{\mathrm{age}}\gtrsim100$\,Myr in this sample on the low-redshift end at $z\sim9-10$. These are the same objects which also have higher stellar masses mentioned before. This is due to the fact that at redshifts $z\sim9-10$ the Balmer-break still falls into the F444W-band, as already mentioned in \citet{atek23}. These three galaxies in particular show a pronounced color offset between the F356W- and the F444W-band which is caused by the presence of a strong Balmer-break as can be seen in the example presented in the upper panel of Fig.~\ref{fig:balmer_break}. The Balmer-break indicates a significant population of older and redder stars than the young massive stars probed at lower wavelengths. This is in agreement with the findings of e.g. \citet{laporte21a} on the existence of evolved stellar populations at $z\gtrsim9$. There are nonetheless two candidates at $z\sim10$ in our sample which do not show a significant excess in the rest-frame optical F444W-band and thus do not seem to have a pronounced Balmer-break (see example in the lower panel of Fig.~\ref{fig:balmer_break}). These galaxies appear to be young with high SFRs and low stellar masses and in general are in accordance with the rest of our sample at higher redshifts for which we only probe the rest-frame UV emission. On the other hand, the Balmer-break in combination with the Lyman-break represents a strong constraint on a galaxy's redshift whereas the absence of the former in these two candidates might instead indicate that we are in fact looking at red low-redshift galaxies as will be discussed in detail in section~\ref{sec:lowz_solutions}.

Finally, while the photometric redshifts were fixed to the \citet{atek23} results with a Gaussian prior in our fit with \texttt{BEAGLE} (section~\ref{sec:BEAGLE-setup}) and therefore broadly agree with them, we find marginally different values for $z_{\mathrm{phot}}$ and its uncertainties and will use these updated values from here on.

\section{Parameter relations} \label{sec:relations}
With the parameters derived from our SED-fitting procedure in section~\ref{sec:SED-fit}, we are now able to place some of these parameters into context and derive the first mass-to-light ratios and mass-SFR relations of lensed galaxies at $z\sim9-16$ in sections~\ref{sec:mass-light} and~\ref{sec:mass-sfr}.

\subsection{Mass-to-light ratios} \label{sec:mass-light}

\begin{figure}
    \centering
    \includegraphics[width=\columnwidth, keepaspectratio=true]{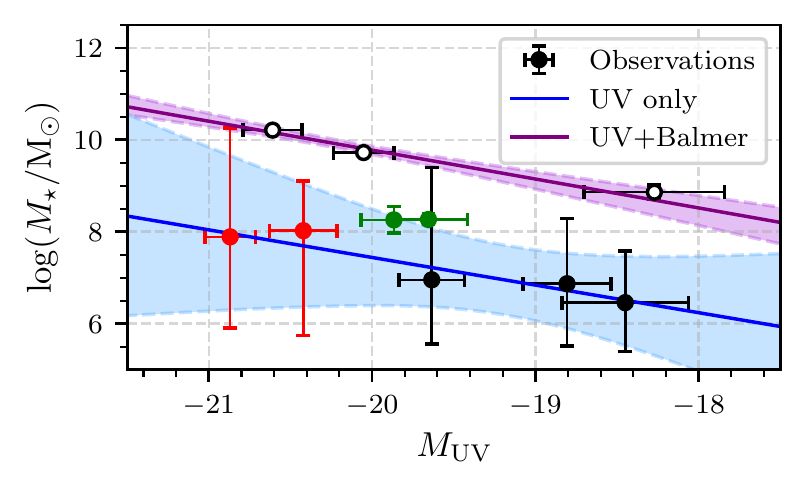}
    \caption{$M_{\star}-M_{\mathrm{UV}}$-relation for our sample of $z\sim9-16$ galaxies detected behind SMACS0723. Black dots represent galaxies at $z\sim9-11$, green dots galaxies at $z\sim12$ and red dots galaxies at $z\sim15-16$ (values given in Tab.~\ref{tab:galay_parameters}). The open black dots represent the three objects at $z\sim9-10$ that show pronounced Balmer-breaks in the F444W-band. Our linear fits are shown with their $1\sigma$-ranges in blue and purple respectively.}
    \label{fig:M-L_ratio}
\end{figure}

The total rest-frame UV luminosity $M_{\mathrm{UV}}$ of each galaxy in our sample is computed as the absolute AB magnitude in the band that contains 1500\,\AA\ at the galaxy's photometric redshift and using the magnification $\mu$ reported in Tab.~\ref{tab:SL}. The UV luminosities are shown in the last column of Tab.~\ref{tab:galay_parameters}. All of our galaxies are UV-bright with luminosities $M_{\mathrm{UV}}\lesssim-18.5$. This is however not surprising since we do not have any extremely magnified objects in this sample (cf. Tab.~\ref{tab:SL}) which means that we miss intrinsically faint sources too faint to be detected without large gravitational magnifications in this sample. Note that this might also be due to selection effects in these JWST observations as discussed in \citet{mason22}.

With both UV luminosity and stellar mass in hand, we are now able to compute the mass-to-light ratios of our $z\sim9-16$ galaxies which are shown in Fig.~\ref{fig:M-L_ratio}. We do not find any significant evolution with redshift apart from the fact that the three $z\sim9-10$ sources with Balmer-breaks have significantly higher stellar masses as already discussed in section~\ref{sec:SED-fit_results} (however also cf. section~\ref{sec:BEAGLE_limits}). These galaxies, shown as open circles in Fig.~\ref{fig:M-L_ratio}, seem to form a separate regime in $M_{\star}-M_{\mathrm{UV}}$ space from the galaxies without Balmer breaks. While we expect to underestimate the stellar masses when no rest-frame optical photometry is available (cf. section~\ref{sec:BEAGLE_limits}), the $z\sim9-10$ objects in our sample that do not show any significant Balmer-breaks in their photometry align well with the higher redshift galaxies for which we only have UV photometry, suggesting that this lower-mass population with lower mass-to-light ratios is genuine. Note that our two $z\sim12$ candidates have stellar masses $M_{\star}\sim10^{8.2}\,\mathrm{M}_{\odot}$ which places them roughly between the two stellar mass populations but well withing the uncertainty range of the lower-mass population. Our very small sample size however does not allow us to draw conclusions about a redshift evolution of stellar mass at this stage.

We fit the $M_{\star}-M_{\mathrm{UV}}$-relation with a linear function using 100 Monte-Carlo Markov Chains (MCMC) of $10^4$ steps each with the \texttt{emcee} package \citep[][]{foreman-mackey13}. The best-fit relations are

\begin{equation} \label{eq:m-l_low}
    \log\left(\frac{M_{\star}}{\mathrm{M}_{\odot}}\right)=(-0.6\pm0.9)(M_{\mathrm{UV}}+20)-(0.6\pm1.0)+8
\end{equation}

for the majority of our sample (blue line in Fig.~\ref{fig:M-L_ratio}) and

\begin{equation} \label{eq:m-l_balmer}
    \log\left(\frac{M_{\star}}{\mathrm{M}_{\odot}}\right)=(-0.6\pm0.2)(M_{\mathrm{UV}}+20)+(1.8\pm0.1)+8
\end{equation}

for the three Balmer-break sources (purple line in Fig.~\ref{fig:M-L_ratio}). Note that we exclude the two $z\sim12$ candidates from the fits since their low stellar mass errors would give them too much relative weight and falsify the $M_{\star}-M_{\mathrm{UV}}$-fit. Both slopes are similar and relatively shallow, at least compared to what has been measured at lower redshifts $z\sim6-9$ \citep[e.g.][]{song16,bhatawdekar19,kikuchihara20,furtak21}. Note however that these relations were also measured for fainter galaxies than our sample, at lower redshifts and with rest-frame optical photometry to constrain eventual evolved stellar populations.

\subsection{SFR-stellar mass relations} \label{sec:mass-sfr}

\begin{figure*}
     \centering
     \hspace{-6mm}
     \begin{subfigure}[b]{1\columnwidth}
         \centering
         \includegraphics[width=1.15\columnwidth]{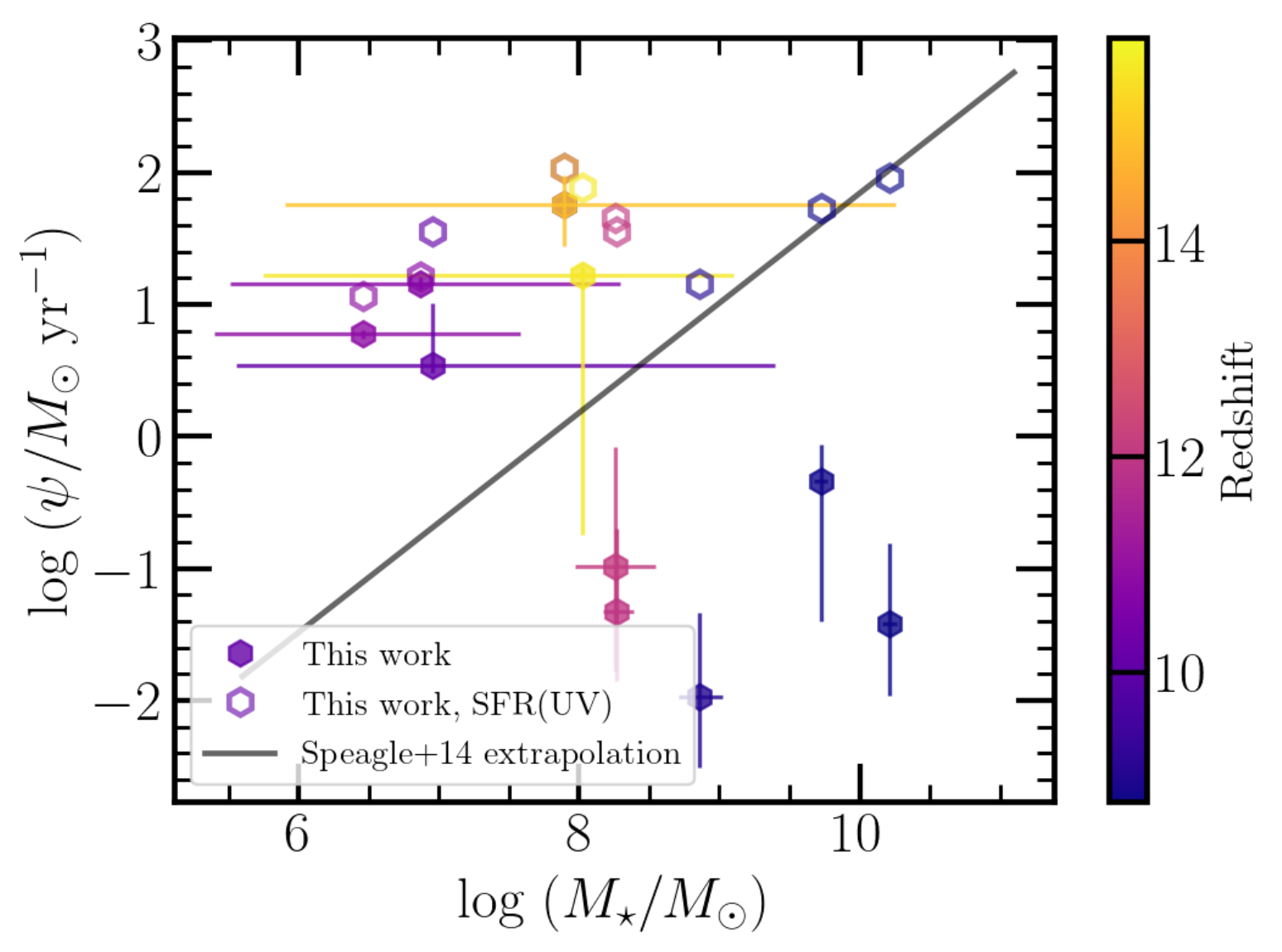}
        %  \caption{}
        %  \label{fig:y equals x}
     \end{subfigure}
     \hspace{2mm}
     \begin{subfigure}[b]{1\columnwidth}
         \centering
         \includegraphics[width=1.1\columnwidth]{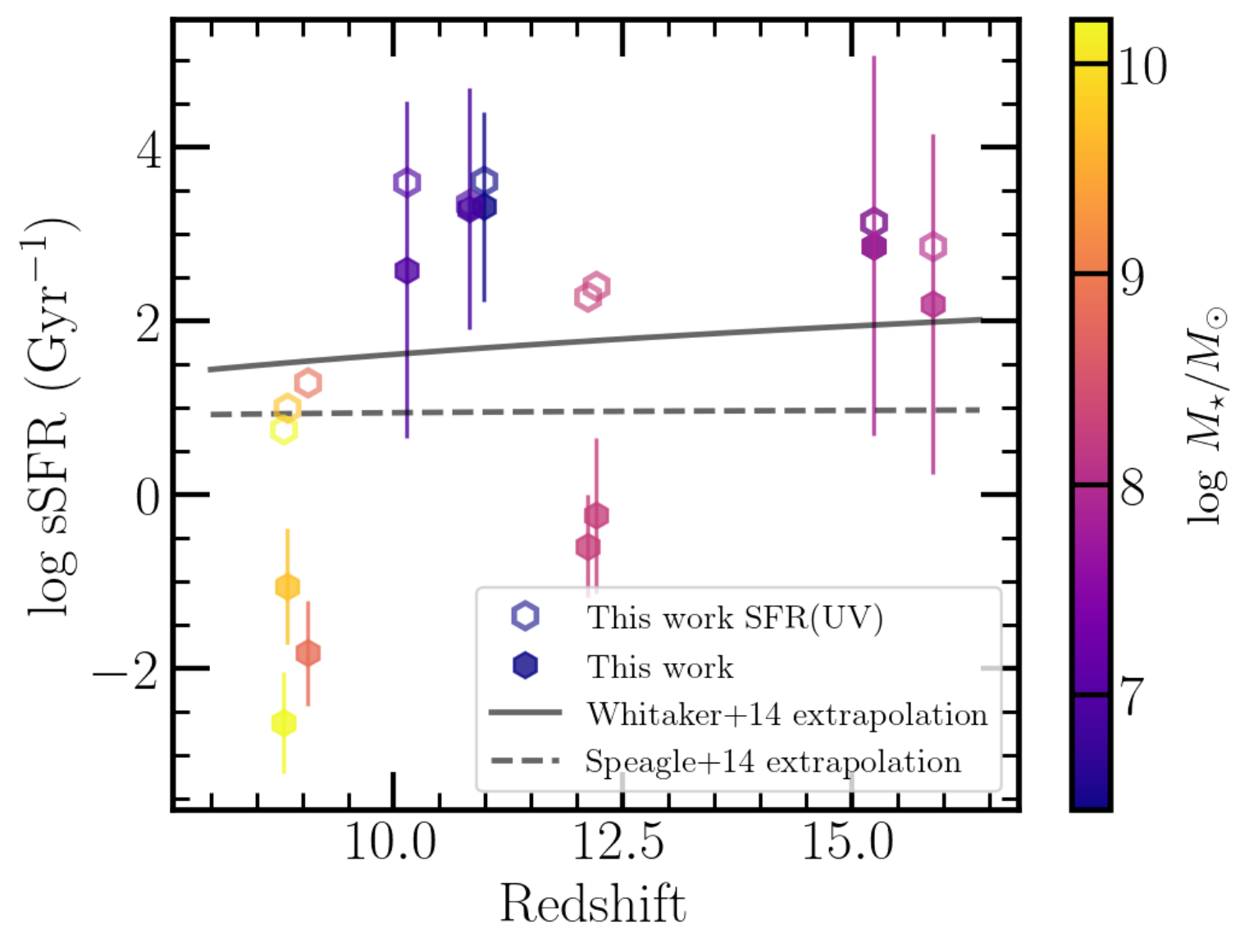}
        %  \caption{}
        %  \label{fig:three sin x}
     \end{subfigure}
        \caption[]{
        SFR as a function of stellar mass and sSFR as a function of redshift for the $z\sim9-16$ galaxy candidates in this study. {\it Left:} SFR as a function of stellar mass color-coded by the photometric redshift. The black line shows the main sequence, the redshift-evolution of which is parametrized in \citet{speagle14} and extrapolated out to $z=13$ in this study. 
        % The majority of our galaxies have elevated SFRs compared to the main sequence. 
        {\it Right:} sSFR as a function of redshift color-coded by the stellar mass. The black curve shows an extrapolation of the sSFR-$z$ evolution as parametrized by \citet{whitaker14} for $\log(M_{\star}/\mathrm{M}_{\odot})\sim9.3$, using SFRs derived from $\beta$-corrected UV luminosities. 
        % Except for the most massive ones, the galaxies in our sample show an elevated sSFR compared to the sSFR of the main sequence roughly constantly over the $z\sim9-16$ range. I
        In both panels, the solid hexagons show the SFRs obtained from the SED-fits with \texttt{BEAGLE} reported in Tab.~\ref{tab:galay_parameters} while the empty hexagons show the SFRs obtained from the UV luminosities using the \citet{kennicutt12} relation.
        }
        \label{fig:SFR-Ms-z}
\end{figure*}

The SFR of galaxies has been established to have a tight correlation with the stellar mass, forming the main sequence (MS) of star-formation \citep{daddi07,noeske07,elbaz07}. The MS has been shown to be in place since high redshifts, and its evolution with redshift has been parametrized out to $z\sim6$ \citep[e.g.,][]{speagle14}. However, the MS is exceptionally difficult to establish at high redshifts due to various systematics and selection effects in compiling representative galaxy samples at such high redshifts \citep[e.g.][]{grazian15,foerster-schreiber20,furtak21}. Despite the unknown territory that is the $z\gtrsim10$ Universe just starting to be revealed by JWST, we analyse the SFR-$M_{\star}$ relation of our sample and compare with extrapolations of the relations calibrated at $z<6$.

Fig.~\ref{fig:SFR-Ms-z} shows the SFR-$M_{\star}$ relation (left-hand panel) and the specific SFR (sSFR;  $\psi_{s}=\psi M_{\star}^{-1}$) for our $z\sim9-16$ sample. The filled hexagons use the SFR derived from the \texttt{BEAGLE} SED-fits reported in Tab.~\ref{tab:galay_parameters}, while the empty hexagons use an SFR estimate derived from the UV luminosity following the \citet{kennicutt12} calibration. The points are color-coded according to their redshift or stellar mass in the left- and right-hand panels respectively. We compare these relations with a parametrized form of the redshift-evolution of the SFR-$M_{\star}$ and sSFR-$z$ relations. We use the SFR-$M_{\star}$ MS vs. $z$ parametrization of \citet{speagle14} that has been established in a consistent analysis of a compilation of 25 studies from the literature. It has been calibrated to $z=6$ but we extrapolate it to $z=13$ and to lower masses for comparison purposes. This is shown in the black solid line in the left-hand panel of Fig.~\ref{fig:SFR-Ms-z}. For the sSFR vs. $z$ relation, we in addition use the parametrization measured by \citet{whitaker14} for $\log(M_{\star}/\mathrm{M}_{\odot})\sim9.3$, using SFRs derived from $\beta$-corrected UV luminosities on complete low-mass samples from $0.5<z<2.5$. This is shown as the black solid line in the right-hand panel of Fig.~\ref{fig:M-L_ratio}.

Our galaxy sample at $z\sim9-16$ shows elevated UV SFR and sSFR compared to the extrapolations from \cite{speagle14} and \cite{whitaker14}. The UV sSFR is of order $\psi_s\gtrsim100\,\mathrm{Gyr}^{-1}$ for the majority of sources, which indicates a very high of star-formation activity. The SFRs estimated from the \texttt{BEAGLE} SED-fits show a larger scatter, and are lower than the UV-inferred ones, especially for the most massive galaxies, which fall below the MS. These are the $z\sim9-10$ galaxies with significant Balmer-break detections and the $z\sim12$ galaxies already mentioned in sections~\ref{sec:SED-fit_results} and~\ref{sec:mass-light}. However, they still fall onto the MS extrapolations and above when considering their UV-inferred SFRs. Overall, the high sSFRs coupled with the relatively young ages measured in our sample indicate that these UV-bright galaxies are observed during a star-bursting episode. This is in agreement with \citet{whitler23} who suggest that the observed number density of bright galaxies at $z>12$ can be explained if they are being observed during a burst of star formation. Their conclusion is based on the estimation of young stellar populations ($\sim30$\,Myr) of bright $z\sim8-11$ galaxies, whose $z\sim15$ progenitors would be relatively faint. This would imply a rapid decline of the number densities of bright galaxies between $z\sim8-11$ and $z\sim15$ which is in tension with the first results from JWST of little-to-no evolution of the large number densities of bright galaxies \citep[e.g.][]{castellano22,naidu22b,atek23}. Observing the bright $z>12$ galaxies during an episode of a star-burst would alleviate this tension. The bursty sSFRs and relatively young ages of our candidates are consistent with this scenario.

Our results are also consistent with the theoretical model of \citet{mason22} which suggests that observed $z\gtrsim10$ galaxies are predominantly extremely star-forming, with young ages, fast formation timescales and high $M_{\rm UV}-M_{\rm halo}$ ratios. This requires high star-formation efficiencies, where a high fraction of the available gas within the dark matter (DM) halo is converted to stars. The elevated star-formation efficiency is also required to explain the apparent observed number-densities of bright galaxies. These, although high compared to current theoretical predictions, are still below the upper limit of the UV luminosity function imposed by a star-formation efficiency of unity \citep{mason22}.

\section{Discussion} \label{sec:discussion}
We have presented our results on the physical parameters of the first lensed $z\sim9-16$ galaxies observed with JWST in SMACS0723 and now discuss these results regarding several issues in this section. First, we use the Balmer-break detections and non-detections to determine if we are missing a significant fraction of galaxies in section~\ref{sec:duty-cycles}. Our analysis is subject to certain limitations which we will discuss in section~\ref{sec:BEAGLE_limits}, in particular regarding the higher redshift objects in our sample. We then furthermore discuss the possibilities of low-redshift interlopers falsely identified as high-redshift sources in section~\ref{sec:lowz_solutions}. Finally, we briefly discuss the implications of our findings for the bigger picture of structure formation in the early Universe in section~\ref{sec:DM}.

\subsection{Star-formation duty cycles in $z\sim9-10$ galaxies} \label{sec:duty-cycles}
As can be seen in Tab.~\ref{tab:galay_parameters} and was pointed out in section~\ref{sec:mass-light}, our galaxy sample is comprised of very bright galaxies with $M_{\mathrm{UV}}\lesssim-19$ which indicates them to be undergoing phases of intense star-formation at the time of observation as discussed in section~\ref{sec:mass-sfr}. It is indeed expected of high-redshift galaxies to not have continuous SFHs but rather episodic star-bursting phases recurring over short duty cycles \citep[e.g.][]{stark09}, in particular in the low-mass and compact dwarf galaxy regime \citep{atek22a}. In this scenario, evolved stellar populations are eventually built-up by each consecutive star-burst. At low redshifts, star-bursts have been found to last for one to a few dynamical timescales \citep[e.g.][]{lehnert96,kennicutt98}.

The fact that we appear to observe galaxies both with and without Balmer-breaks at $z\sim9-10$ therefore enables us to perform the following thought-experiment: The selection window for the $z\sim9-11$ galaxies in our sample \citep{atek23} corresponds to a time $t_{\mathrm{obs}}\sim200$\,Myr during which these galaxies can be observed. Our Balmer-break galaxies have ages up to $10^{8.3}$\,yr which results in $t_{\mathrm{age}}/t_{\mathrm{obs}}\sim1$. Interestingly, $t_{\mathrm{age}}/t_{\mathrm{obs}}\lesssim0.5$ for the two $z\sim9-10$ objects without Balmer-break detections which is relatively close to the ratio of Balmer-break to total $z\sim9-10$ objects of $0.4$. This indicates that we might not necessarily be missing a significant population of dark galaxies but are instead simply looking at episodic star-bursting galaxies at various stages within their star-bursting duty cycles: older galaxies that already have built-up a population of red stars and are currently in between two star-bursts and younger galaxies that are currently within their first or first few duty cycles of star-formation.

Interestingly, the Balmer-break galaxies in our sample show a significant difference between their SED-fitting and UV luminosity inferred SFRs (cf. Fig.~\ref{fig:SFR-Ms-z}). These two SFR measurements typically probe different time scales, $\sim10$\,Myr for the \texttt{BEAGLE}-inferred SFRs (cf. section~\ref{sec:BEAGLE-setup}) and averaged over $\sim100-200$\,Myr for UV luminosity inferred SFRs \citep[e.g.][]{leitherer99,hao11,kennicutt12,calzetti13}. For our Balmer break objects, where the SED-fitting SFRs are significantly lower than the UV SFRs, this discrepancy could therefore indicate that we might be observing these galaxies within 100-200\,Myr after a strong star-bursting episode \citep[e.g.][]{weisz12,dominguez15,emami19} which would be in accordance with our hypothesis from above that we might be looking at star-bursting galaxies at various stages within or between star-bursting episodes. Note that since we are here talking about only 5 galaxies for which the rest-frame optical photometry is available, we do of course not have enough statistics to support this scenario and will require both much larger sample sizes over different observation fields to obtain real statistics and spectroscopic observations to robustly measure the current SFRs from the nebular emission lines.

\subsection{The limits of our SED-analysis} \label{sec:BEAGLE_limits}

\begin{figure}
    \centering
    \includegraphics[width=\columnwidth, keepaspectratio=true]{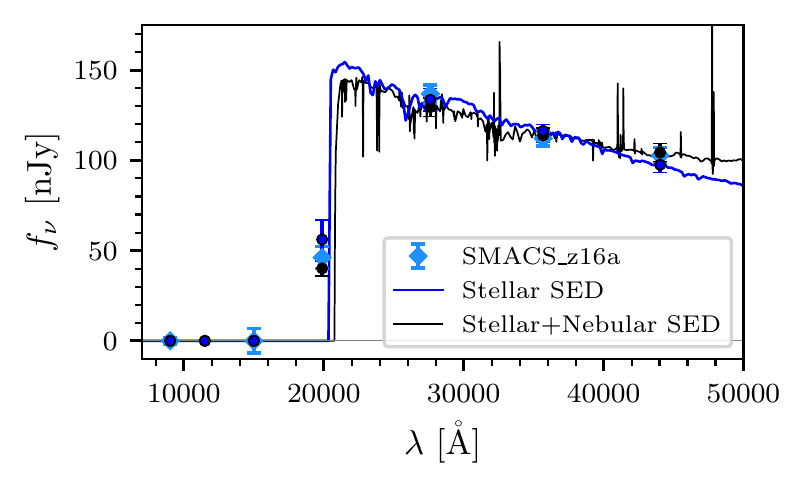}
    \includegraphics[width=\columnwidth, keepaspectratio=true]{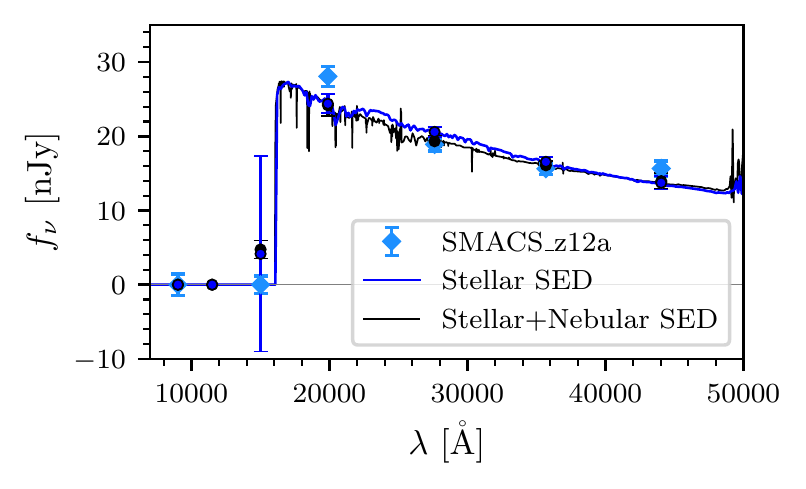}
    \caption{Best-fit \texttt{BEAGLE} SEDs of SMACS\_z16a (upper panel) and SMACS\_z12a (lower panel) using stellar and nebular emission templates (black) and stellar templates only (blue). The blue diamonds represent the observed photometry. The stellar continuum-only SEDs seem to provide similar fits to these galaxies as the SED including nebular emission despite their extremely blue UV-slopes, $\beta\simeq-2.6$ and $\beta\simeq-2.8$ respectively.}
    \label{fig:continuum-fit}
\end{figure}

This study heavily relies on the Bayesian SED-fitting tool \texttt{BEAGLE}, the effectiveness and flexibility of which has been proven in numerous galaxy studies \citep[e.g.][]{chevallard16,plat19,endsley21,furtak21,topping22}. However, we nonetheless estimate our SED-fitting analysis of this $z\sim9-16$ galaxy sample to be prone to three major limitations detailed in the following.

First, as already mentioned in sections~\ref{sec:SED-fit_results} and~\ref{sec:mass-light}, we are limited to rest-frame UV photometry for all of our galaxies beyond $z>10$. Since the rest-frame UV wavelengths only probe the short-lived young and massive stars of a galaxy which do not necessarily make out the bulk of its mass (though cf. section \ref{sec:duty-cycles}), stellar masses can only accurately be derived from the UV continuum if a galaxy is $\lesssim10$\,Myr old (i.e. the average life-time of massive stars). Beyond that, accurate derivation of the stellar mass requires a continuum detection red-ward of the Balmer-break. This is further corroborated by the fact that the stellar masses of our three galaxies with Balmer-break detections have very small uncertainties whereas the uncertainties on the stellar masses of many other galaxies in our sample are very large (cf. Tab.~\ref{tab:galay_parameters}). More quantitatively, it has been shown that SED-fitting UV photometry only can underestimate stellar masses by up to 0.6\,dex \citep[][]{furtak21}. Interestingly, that is about the order of magnitude that our results are offset from the $M_{\star}$-SFR main sequence as shown in Fig.~\ref{fig:SFR-Ms-z} and discussed in section~\ref{sec:mass-sfr}. In order to determine if galaxies at $z\gtrsim10$ do have an evolved stellar population, which is predicted by recent simulations \citep[][]{mason22}, the JWST/NIRCam imaging at wavelengths $\lambda\leq5\,\mu$m will need to be complemented by deep imaging with the \textit{Mid-Infrared Instrument} \citep[MIRI;][]{bouchet12,rieke15}, also aboard the JWST, which has also already proven its unprecedented sensitivity at wavelengths $\lambda>5\,\mu$m \citep[][]{ling22}. This will allow to probe the rest-frame optical emission of these very high redshift galaxies.

The next limitation of our analysis is that the fiducial templates used by \texttt{BEAGLE} are ionization bounded, i.e. assume a Lyman-continuum (LyC) escape fraction of $f_{\mathrm{esc}}=0$. The escape fraction of LyC however has a significant impact on the shape of the SED through the nebular emission: It weakens both nebular line and continuum emission which makes the SED bluer and the UV-slope steeper \citep[e.g.][]{zackrisson13,zackrisson17,plat19}. Indeed, observations of known LyC leakers at low redshifts clearly show a correlation between UV-slope and LyC escape fraction \citep[e.g.][]{chisholm22}. This is a particularly important effect to take into account in highly star-forming primeval galaxies in the early Universe as it could lead to difficulties reproducing the extremely blue UV-slopes of some of some JWST-detected $z\gtrsim10$ galaxies. In our sample in particular, the bluest UV-slopes are measured for the four highest-redshift candidates at $z\sim12-16$ \citep[down to $\beta\simeq-2.8$;][]{atek23}. To verify if the SED-fits performed in section~\ref{sec:SED-fit} hit this ionization boundary limit, we also fit the photometry of our candidates in a separate \texttt{BEAGLE} run using only stellar continuum templates and show the two bluest galaxies, SMACS\_z16a and SMACS\_z12a, in Fig.~\ref{fig:continuum-fit}. As can be seen from the blue lines in Fig.~\ref{fig:continuum-fit}, this run without nebular continuum provides a similar fit to the initial run that includes nebular emission which proves the validity our SED-analysis. Nonetheless, this possible issue needs to be kept in mind in future high-redshift studies with JWST. In order to derive accurate physical parameters of high-redshift galaxies through SED-fitting, future studies will therefore need to include templates that allow for $f_{\mathrm{esc}}>0$, such as e.g. those by \citet{plat19}, as was already done for some JWST sources in \citet{topping22}. Extremely blue UV-slopes such as presented by our highest-redshift candidates at $z\gtrsim12$ ($\beta\lesssim-2.5$) are rarely observed at low redshifts and often in galaxies with peculiar properties such as exotic nebular emission spectra or LyC leakage \citep[e.g.][]{furtak22,chisholm22}. These might however be more common features at very high redshifts $z\gtrsim12$. Indeed, while \citet{cullen22} find the average UV-slopes at $z\sim8-15$ to not be significantly bluer than the slopes at $z\lesssim8$, individual galaxies at very high redshifts up to $z\sim16$ seem to strongly tend towards extreme UV-slopes down to $\beta\sim-3$ \citep{atek23,topping22}.

Finally, another possible issue in our analysis is the metallicity since the templates used in \texttt{BEAGLE} are limited to metallicities $Z\geq0.01\,\mathrm{Z}_{\odot}$ \citep{chevallard16}. As can be seen in Tab.~\ref{tab:galay_parameters}, we find metallicities in the range $Z\sim0.01-0.1\,\mathrm{Z}_{\odot}$ for the majority of our galaxies which concurs with the findings of \citet{topping22}. Since galaxies at $z\gtrsim15$ represent the first luminous structures formed in the Universe, there is a distinct possibility that in particular our highest-redshift candidates at $z\sim16$ are in fact systems that contain significant amounts pristine Population~III (Pop.~III) stars which would have metallicities approaching $Z=0$. If this were the case, we would however not be able to measure it because: (i) the \texttt{BEAGLE} templates do not extend that far in metallicity space and (ii) the rest-frame UV photometry is not sensitive to stellar metallicity \citep[e.g.][]{furtak21}. Note also that simulations have found Pop.~III dominated galaxies to be too faint to be detected with JWST \citep[][]{riaz22}. On the other hand, simulations also suggest that the very first stars would rapidly enrich their surrounding medium in metals and thus make galaxies with significant metallicities possible even at very high redshifts \citep[][]{sanati22}. Accurate measurements of the metallicities of these kinds of galaxies will require ultra-deep spectroscopy with JWST's \textit{Near-Infrared Spectrograph} \citep[NIRSpec;][]{jakobsen22} to search for spectroscopic signatures of Pop.~III stars in the highest redshift galaxies \citep[e.g.][]{cassata13,berzin21,katz22a}.

\subsection{Possible contamination by low-redshift interlopers} \label{sec:lowz_solutions}

\begin{figure}
    \centering
    \includegraphics[width=\columnwidth, keepaspectratio=true]{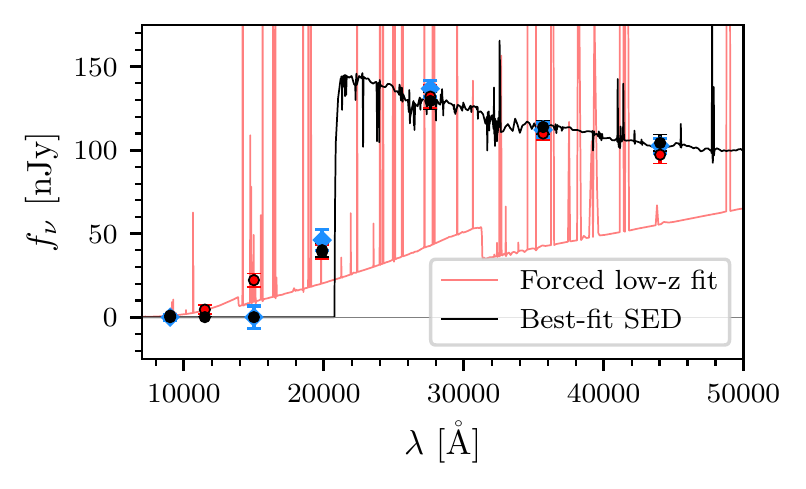}
    \caption{Best-fit \texttt{BEAGLE} high- (black) and forced low-redshift (red) SED for the $z\sim16$ candidate SMACS\_z16a also shown in Figs.~\ref{fig:z16a_fit} and~\ref{fig:continuum-fit}. The blue solid black and red dots represent the predicted fluxes of the corresponding SED. Note that the forced low-redshift SED-fit has parameters dredging-up against the range of allowed values (e.g. stellar age) and thus represents a bad fit. The nebular emission lines of the low-redshift fit have EWs up to $\sim2500$\,\AA.}
    \label{fig:lowz_nebular-fit}
\end{figure}

\begin{figure}
    \centering
    \includegraphics[width=\columnwidth, keepaspectratio=true]{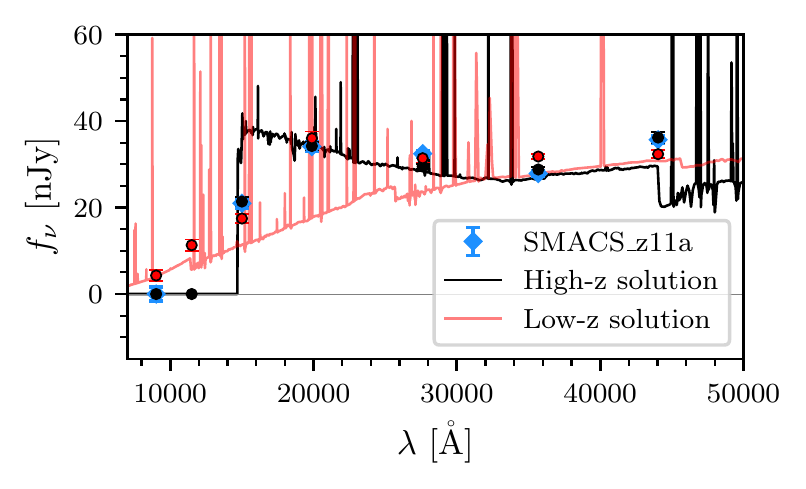}
    \caption{Best-fit high- (black) and forced low-redshift (red) SED for the $z\sim11$ candidate SMACS\_z11a. Given the less pronounced UV-slope compared to the bluer galaxies in our sample, the forced low-redshift solution also provides a reasonable fit to the observed photometry.}
    \label{fig:lowz_fit}
\end{figure}

Photometric selection of high-redshift galaxies based on broad-band photometry has always been prone to contamination by low-redshift galaxies and cold stars which, if they are red and faint enough, can mimic the colors and brightness of high-redshift galaxies. For example, studies with HST have shown that dropout selection techniques can reach low-redshift contamination levels of up to 40\,\% \citep[e.g.][]{bouwens11}. We therefore need to take this possibility into account. This is in particular true in the light of recent analyses of the JWST ERO and ERS data which have revealed a class of very dusty star-forming spiral galaxies at low redshifts whose color-signature is indistinguishable from that of $z\gtrsim10$ galaxies in the JWST imaging \citep[][]{fudamoto22,nelson22,zavala22,naidu22c,glazebrook22}. Indeed, millimeter observations of the $z\sim17$ candidate detected in the CEERS field by \citet{donnan23} have already revealed it to possibly be a $z\sim5$ dusty star-forming galaxy (DSFG) instead \citep{zavala22,naidu22c}. Note that this could in part explain the disk-like morphologies measured for JWST high-redshift candidates \citep[e.g.][]{naidu22b}, including our sample \citep[][]{atek23}.

In an attempt to assess the level of low-redshift contamination of our sample we conduct additional SED fits with both \texttt{BEAGLE} and \texttt{EAZY} \citep[][]{brammer08} in which we force the photometric redshifts to low values $z<9$ and allow older and more dusty galaxies as would be expected at low redshifts. We find that the best-fitting forced low-redshift SEDs in general provide fits of lower significance than the open redshift ones performed in \citet{atek23} which clearly favor the high-redshift solutions. They are in particular incapable of reproducing the UV-slopes of the bluest of our candidates, even with extremely strong nebular emission lines (equivalent widths up to $\mathrm{EW}_0\sim2500$\,\AA), as illustrated by the example shown in Fig.~\ref{fig:lowz_nebular-fit}. The forced low-redshift fits in these cases in particular favor stellar ages that but against the lower boundary ($\log(t_{\mathrm{age}}/\mathrm{yr})\lesssim6.3$ at $3\sigma$), i.e. the dynamical time as explained in section~\ref{sec:BEAGLE-setup}, and prefers solutions with significant extinctions, $A_V\sim2$. Since galaxies younger than their dynamical time are deemed not physical (cf. sections~\ref{sec:priors} and~\ref{sec:duty-cycles}), we can discard the forced low-redshift fits for these sources. We therefore estimate that the sources in our sample with the steepest UV-slopes $\beta\lesssim-2.5$, which include our highest-redshift candidate SMACS\_z16a at $z\sim16$ and the two $z\sim12$ candidates, are most probably robust high-redshift galaxies. In addition, both SED-fitting codes are incapable to find a low-redshift solution for the two $z\sim9-10$ objects that have Balmer-break detections (cf. section~\ref{sec:SED-fit_results}) which suggests that the combination of the Lyman- and Balmer-breaks is a robust probe of photometric redshift. For the remaining three candidates in our sample, this is more unclear and a low-redshift SED that provides a somewhat reasonable alternative fit can be found as can be seen in the example given in Fig.~\ref{fig:lowz_fit}. For these objects we conclude that the high- and low-redshift solutions are hard to distinguish with the current type of data, i.e. broad-band rest-frame UV photometry.

While all galaxies in our sample passed the rigorous selection process described in \citet{atek23} and have clearly preferred high-redshift photometric redshift estimates, the only way to definitely nail-down their redshifts are of course deep NIRSpec observations to obtain spectroscopic redshifts, as already done for the first time in \citet{roberts-borsani22} and \citet{williams22}. Another useful indicator would be millimeter and sub-millimeter observations with the \textit{Atacama Large Millimeter/sub-millimeter Array} (ALMA) which would probe the rest-frame far infrared (FIR) emission of these galaxies, as already attempted in \citet{fujimoto22}. Direct detections of rest-frame FIR emission lines or dust continuum can provide strong constraints on the dust and would allow distinguishing between a high-redshift galaxy and a low-redshift DSFG as demonstrated in \citet{zavala22}.

\subsection{Implications for scenarios of galaxy formation in DM halos} \label{sec:DM}

\begin{figure}
    \centering
    \includegraphics[width=\columnwidth, keepaspectratio=true]{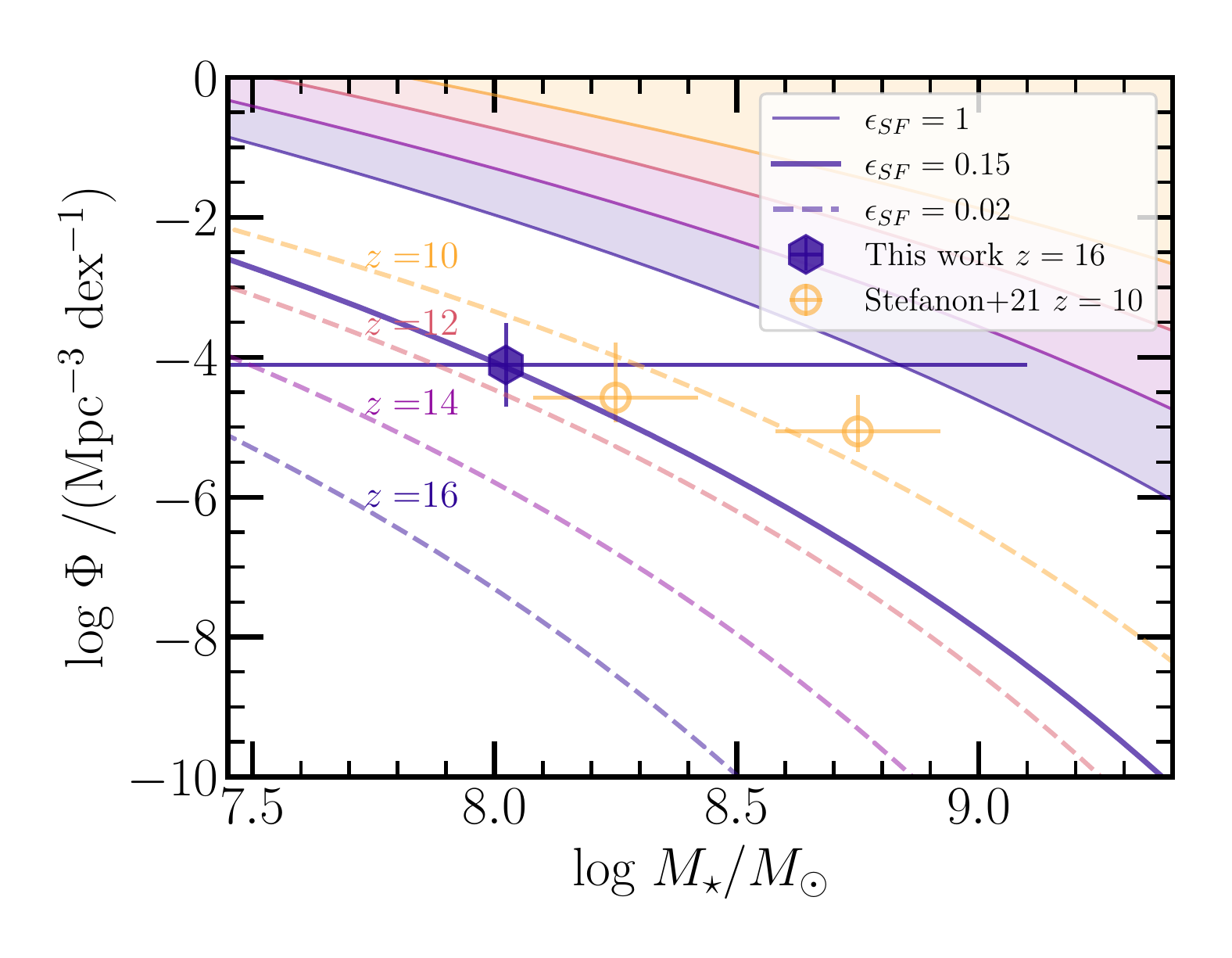}
    \caption{Number density as a function of stellar mass at $z>10$. The hexagon shows the number density of the $z\sim16$ galaxies from this work while the empty circles show the results at $z\sim10$ by \citet{stefanon21}. The thin solid lines with shaded regions mark the upper limits imposed by $100\%$ SFE for $z=10,\, 12, \,14 \, \& \, 16$, color-coded accordingly. The dashed lines show an SFE of $\sim 2 \%$ as estimated by studies at $z<10$. The thick solid line represents the SFE necessary to explain the observed number density of the $z\sim16$ candidates in this study.}
    \label{fig:N_dens}
\end{figure}

One of the most surprising aspects of the numerous reported discoveries of $z\gtrsim10$ galaxies with JWST have been their exceptionally high stellar masses, UV luminosities and number densities. Considerations of the baryon-to-stellar mass conversion in DM halos have already pointed out tensions with what is allowed by theoretical models based on the $\Lambda$CDM paradigm \citep{boylan-kolchin22,naidu22c}. Within this paradigm, galaxies form in DM halos by conversion of the available baryonic gas given by the cosmic baryon fraction $f_{\rm b}\sim0.16$, resulting in a relation between the stellar and the halo mass $M_{\star}=\epsilon_{\rm SF}\,f_{\rm b}\,M_{\rm h}$, as well as between their number densities,

\begin{equation} \label{eq:Phis}
    \Phi_{\star}(M_{\star}, z) = \Phi_{\rm h}(M_{\star}\, f_b^{-1}\, \epsilon_{\rm SF}^{-1},z)
\end{equation} 

\noindent
where $\epsilon_{\rm SF}$ is the efficiency of converting baryons to stars, i.e., the star-formation efficiency (SFE). This relation imposes an upper limit on the maximum stellar mass and stellar mass function that can be formed in a DM halo if the efficiency is $100\%$, i.e. all the available baryons are turned into stars \citep{behroozi18}.

In Fig.~\ref{fig:N_dens}, we put the stellar mass and the number density of the $z\sim16$ galaxies in this work into perspective by comparing with theoretical predictions adopting a similar approach as e.g. \citet{behroozi18,mason22,boylan-kolchin22,naidu22c}. For the theoretical predictions, we use the halo mass function (HMF) from \citet{sheth01} and Eq.~\ref{eq:Phis} to predict the stellar mass number density. The upper limit imposed by $\epsilon_{\rm SF} = 100\%$ for $z=10-16$ are marked by the thin solid lines and shaded regions. For comparison, we show the stellar mass number densities at $z\sim10$ from \citet{stefanon21} estimated from Lyman-break galaxies in  HST and {\it Spitzer Space Telescope} observations. We also over-plot the stellar mass functions predicted by assuming an SFE of $\sim2\%$, which is indicated by numerous studies at $z<10$ \citep[e.g.,][]{behroozi13,tacchella18,stefanon21,shuntov22}. It is worth noting that quantitatively, this comparison is very sensitive to the choice of halo mass function. Nonetheless, qualitatively, the comparisons are sufficiently robust for the purpose of this discussion here.

The number density of our $z\sim16$ candidates indicates $\epsilon_{\rm SF} \sim 0.15$ -- an elevated SFE by a factor of about 7 compared to $z<10$ measurements. This number density is comparable to that at $z\sim10$ assuming $\epsilon_{\rm SF} \sim 0.02$, suggesting no significant evolution of the stellar mass number density at these epochs. This is in tension with findings by previous studies that find no evolution of the SFE \citep[e.g.,][]{mason15,tacchella18}.
One way to reconcile this is with an evolution of the SFE, as indicated by the thick solid line that marks a SFE of about $15\%$ at $z\sim 16$. However, we should be cautious with constraining the SFE in the early Universe from observations such as the ones presented in this study. This is because in addition to the issues discussed in sections~\ref{sec:BEAGLE_limits} and~\ref{sec:lowz_solutions}, there is a scatter in the relation between stellar and halo mass and these galaxies might not necessarily be representative of the whole population. Indeed, they are likely extreme cases of highly star-forming and bright galaxies with negligible dust attenuations \citep[e.g.][]{mason22,ferrara22} as pointed out at multiple occasions in this work. Finally, despite the disagreement with theoretical predictions of the evolution of the number density, our estimates are within the limits imposed by $\Lambda$CDM with $\epsilon_{\rm SF}=1$. 

Note that the number density discussed in this section is of course only a very crude estimate since: (i) we lack statistics with only two $z\sim16$ candidates and (ii) completeness and accurate survey volume that probe all the underlying uncertainties in lensing cluster observations are highly non-trivial to compute and have been shown to have large and complex effects on the derivation of number densities \citep{bouwens17,bouwens22b,atek18,furtak21} which is beyond the scope of this work.

\section{Conclusion} \label{sec:conclusion}
We presented a detailed SED-fitting analysis of the 10 lensed $z\sim9-16$ galaxy candidates detected in the JWST ERO observations of the SL cluster SMACS0723 in \citet{atek23}. This is the first sample of gravitationally lensed galaxies at these extremely high redshifts observed with JWST. Using the \texttt{BEAGLE} tool with priors derived from photometric and morphological measurements and combining all of the SL information currently available, we carefully derived physical parameters of these galaxies and robustly probed their uncertainty space. We then put these results in relation to each other and computed mass-to-light ratios and mass-SFR relations. Finally, we discussed our results regarding SED-fitting limitations and probed the robustness of our sample regarding low-redshift galaxies that can imitate the photometric signatures of high-redshift galaxies. Our main results are the following:

\begin{itemize}
    \item We find the $z\sim9-16$ candidates in our sample to in general have relatively low stellar masses, $M_{\star}\sim10^{6.5}-10^{8.3}\,\mathrm{M}_{\odot}$, young ages $t_{\mathrm{age}}\sim10-100$\,Myr and very low dust attenuations $A_V\sim0.01$ based on their UV-continuum slope.
    \item The detection of strong Balmer-breaks in some of our $z\sim10$ candidates results in significantly higher stellar masses ($M_{\star}\sim10^9-10^{10}\,\mathrm{M}_{\odot}$) and ages ($t_{\mathrm{age}}\gtrsim100$\,Myr) which confirms that evolved stellar populations exist even at redshifts $z\gtrsim10$.
    \item The non-detection of Balmer-breaks in other galaxies at $z\sim9-10$, for which the F444W-band probes rest-frame optical wavelengths, also indicates the existence of a younger and lower-mass population of galaxies at $z\sim9-10$ the properties of which are similar to the rest of our sample at higher redshifts. There is however also a possibility that these objects are low-redshift interlopers.
    \item We find a relatively shallow mass-to-light relation (slopes $\simeq-0.6$) that does not seem to significantly evolve with redshift as the majority of our sources fall onto that relation. The $z\sim9-10$ galaxies with Balmer-break detections for a separate population in $M_{\star}-M_{\mathrm{UV}}$-space but present a very similar mass-to-light relation slope.
    \item The $z\sim9-16$ galaxies in our sample have relatively high sSFRs that lie above the main sequence of star-formation extrapolated out to these high redshifts. Combined with the low stellar masses and young ages in our sample, this indicates that these galaxies are going through an episode of intense star-formation while they are observed. This also explains their low mass-to-light ratios.
    \item Since we are observing rest-frame UV emission only for the galaxies at $z>10$, we are only probing the current episode of star-formation in these galaxies and are perhaps missing older stellar populations. However, the existence of $z\sim9-10$ galaxies without Balmer-breaks in our sample and whose properties align with the UV-inferred ones from the higher-redshift objects, indicates that young galaxies in their first episode of star-formation are also present at these redshifts.
    \item There is no significant evolution of parameters with redshift other than a tendency for the highest-redshift objects ($z\geq12$) to also have the bluest UV-continuum slopes.
    \item Two photometric signatures indicate robust high-redshift solutions: Extremely blue UV-slopes ($\beta\lesssim-2.5$) and Balmer-break detections in rest-frame optical bands in addition to the Lyman-break used for selection. Half of our galaxy sample full-fills at least one of these conditions, including our highest-redshift candidate SMACS\_z16a at $z_{\mathrm{phot}}\simeq15.93_{-0.11}^{+0.11}$ and the two $z\sim12$ candidates. For the remaining objects in our sample, possible low-redshift solutions cannot robustly be ruled out with the data currently available.
    \item A crude estimate of number density of galaxies at $z\sim16$ inferred from our sample is consistent with expected DM halo mass functions if the SFE is higher ($\epsilon_{\mathrm{SF}}\sim 0.15$) than expected at lower redshifts $z\sim10$.
\end{itemize}

Our results in general demonstrate JWST's unprecedented ability to not only detect but also characterize galaxies in the early Universe at $z\gtrsim10$ which foreshadows many more new discoveries to come. Since our sample does not contain any high-magnification sources, likely due to the small critical area of SMACS0723, we only probe the UV-bright galaxy population in this study. Upcoming JWST observations of other SL clusters, for example those  with much larger critical areas than SMACS0723, may therefore yield some highly magnified sources and probe the UV-fainter, lower-mass or less intensely star-forming population of galaxies at $z\gtrsim10$. Since at these redshifts we are observing the rest-frame UV emission with NIRCam, studies such as this one can only robustly probe the currently ongoing episode of star-formation in these galaxies. It is therefore crucial that observations of this nature be complemented by deep MIRI imaging in the future in order to also probe the rest-frame optical emission of these galaxies at $z\gtrsim10$ and better constrain their physical properties. Rest-frame FIR follow-up observations with ALMA will be of great use to distinguish between very high-redshift galaxies and very dusty objects at lower redshifts. Ideally, given the high rate of possible low-redshift contamination, we will need spectroscopic observations with NIRSpec to confirm the high-redshift nature of these galaxies and further constrain their properties. Our highest-redshift candidate, SMACS\_z16a, and the two $z\sim12$ candidates are of particular interest for spectroscopic follow-up observations because of their intriguingly steep UV-slopes, a feature rarely observed at low redshifts, which indicates that their high photometric redshift is genuine. Given these object's properties, we might well be looking at representatives of the very first generations of galaxies formed in the Universe.

\section*{Acknowledgements}
We warmly thank the anonymous referee for their comments and feedback which greatly helped to improve the paper. LF and AZ acknowledge support by Grant No. 2020750 from the United States-Israel Binational Science Foundation (BSF) and Grant No. 2109066 from the United States National Science Foundation (NSF) and support by the Ministry of Science \& Technology, Israel. HA acknowledges support from CNES. JC acknowledges funding from the FirstGalaxies Advanced Grant from the European Research Council (ERC) under the European Union’s Horizon 2020 research and innovation program (Grant agreement No.~789056).

\noindent This study is based on observations obtained with the NASA/ESA/CSA JWST, retrieved from the \texttt{Barbara A. Mikulski Archive for Space Telescopes} (\texttt{MAST}) at the \textit{Space Telescope Science Institute} (STScI). STScI is operated by the Association of Universities for Research in Astronomy, Inc. under NASA contract NAS 5-26555. This study has made use of the \texttt{CANDIDE} Cluster at the \textit{Institut d'Astrophysique de Paris} (IAP), made possible by grants from the PNCG and the region of Île de France through the program DIM-ACAV+. This research made use of \texttt{Astropy},\footnote{\url{http://www.astropy.org}} a community-developed core Python package for Astronomy \citep{astropy13,astropy18} as well as the packages \texttt{NumPy} \citep{vanderwalt11}, \texttt{SciPy} \citep{virtanen20} and \texttt{Matplotlib} \citep{hunter07}. This research also made use of {\tt SourceXtractor++}\footnote{\url{https://github.com/astrorama/SourceXtractorPlusPlus}}, an open source software package developed for the \textit{Euclid} satellite project.
 
\section*{Data Availability}
The data underlying this article are publicly available on the \texttt{Barbara A. Mikulski Archive for Space Telescopes}\footnote{\url{https://archive.stsci.edu/}} (\texttt{MAST}), under program ID 2736.

\bibliographystyle{mnras}
\bibliography{references}

\begin{thebibliography}{}
\makeatletter
\relax
\def\mn@urlcharsother{\let\do\@makeother \do\$\do\&\do\#\do\^\do\_\do\%\do\~}
\def\mn@doi{\begingroup\mn@urlcharsother \@ifnextchar [ {\mn@doi@}
  {\mn@doi@[]}}
\def\mn@doi@[#1]#2{\def\@tempa{#1}\ifx\@tempa\@empty \href
  {http://dx.doi.org/#2} {doi:#2}\else \href {http://dx.doi.org/#2} {#1}\fi
  \endgroup}
\def\mn@eprint#1#2{\mn@eprint@#1:#2::\@nil}
\def\mn@eprint@arXiv#1{\href {http://arxiv.org/abs/#1} {{\tt arXiv:#1}}}
\def\mn@eprint@dblp#1{\href {http://dblp.uni-trier.de/rec/bibtex/#1.xml}
  {dblp:#1}}
\def\mn@eprint@#1:#2:#3:#4\@nil{\def\@tempa {#1}\def\@tempb {#2}\def\@tempc
  {#3}\ifx \@tempc \@empty \let \@tempc \@tempb \let \@tempb \@tempa \fi \ifx
  \@tempb \@empty \def\@tempb {arXiv}\fi \@ifundefined
  {mn@eprint@\@tempb}{\@tempb:\@tempc}{\expandafter \expandafter \csname
  mn@eprint@\@tempb\endcsname \expandafter{\@tempc}}}

\bibitem[\protect\citeauthoryear{{Adams} et~al.,}{{Adams}
  et~al.}{2023}]{adams23}
{Adams} N.~J.,  et~al., 2023, \mn@doi [\mnras] {10.1093/mnras/stac3347}, \href
  {https://ui.adsabs.harvard.edu/abs/2023MNRAS.518.4755A} {518, 4755}

\bibitem[\protect\citeauthoryear{{Astropy Collaboration} et~al.,}{{Astropy
  Collaboration} et~al.}{2013}]{astropy13}
{Astropy Collaboration} et~al., 2013, \mn@doi [\aap]
  {10.1051/0004-6361/201322068}, \href
  {http://adsabs.harvard.edu/abs/2013A%26A...558A..33A} {558, A33}

\bibitem[\protect\citeauthoryear{{Atek}, {Richard}, {Kneib}  \&
  {Schaerer}}{{Atek} et~al.}{2018}]{atek18}
{Atek} H.,  {Richard} J.,  {Kneib} J.-P.,   {Schaerer} D.,  2018, \mn@doi
  [\mnras] {10.1093/mnras/sty1820}, \href
  {https://ui.adsabs.harvard.edu/abs/2018MNRAS.479.5184A} {479, 5184}

\bibitem[\protect\citeauthoryear{{Atek}, {Furtak}, {Oesch}, {van Dokkum},
  {Reddy}, {Contini}, {Illingworth}  \& {Wilkins}}{{Atek}
  et~al.}{2022}]{atek22a}
{Atek} H.,  {Furtak} L.~J.,  {Oesch} P.,  {van Dokkum} P.,  {Reddy} N.,
  {Contini} T.,  {Illingworth} G.,   {Wilkins} S.,  2022, \mn@doi [\mnras]
  {10.1093/mnras/stac360}, \href
  {https://ui.adsabs.harvard.edu/abs/2022MNRAS.511.4464A} {511, 4464}

\bibitem[\protect\citeauthoryear{{Atek} et~al.,}{{Atek} et~al.}{2023}]{atek23}
{Atek} H.,  et~al., 2023, \mn@doi [\mnras] {10.1093/mnras/stac3144}, \href
  {https://ui.adsabs.harvard.edu/abs/2023MNRAS.519.1201A} {519, 1201}

\bibitem[\protect\citeauthoryear{{Bagley} et~al.,}{{Bagley}
  et~al.}{2022}]{bagley22}
{Bagley} M.~B.,  et~al., 2022, arXiv e-prints, \href
  {https://ui.adsabs.harvard.edu/abs/2022arXiv221102495B} {p. arXiv:2211.02495}

\bibitem[\protect\citeauthoryear{{Barrufet} et~al.,}{{Barrufet}
  et~al.}{2022}]{barrufet22}
{Barrufet} L.,  et~al., 2022, arXiv e-prints, \href
  {https://ui.adsabs.harvard.edu/abs/2022arXiv220714733B} {p. arXiv:2207.14733}

\bibitem[\protect\citeauthoryear{{Behroozi} \& {Silk}}{{Behroozi} \&
  {Silk}}{2018}]{behroozi18}
{Behroozi} P.,  {Silk} J.,  2018, \mn@doi [\mnras] {10.1093/mnras/sty945},
  \href {https://ui.adsabs.harvard.edu/abs/2018MNRAS.477.5382B} {477, 5382}

\bibitem[\protect\citeauthoryear{{Behroozi}, {Wechsler}  \&
  {Conroy}}{{Behroozi} et~al.}{2013}]{behroozi13}
{Behroozi} P.~S.,  {Wechsler} R.~H.,   {Conroy} C.,  2013, \mn@doi [\apj]
  {10.1088/0004-637X/770/1/57}, \href
  {http://adsabs.harvard.edu/abs/2013ApJ...770...57B} {770, 57}

\bibitem[\protect\citeauthoryear{{Bertin}, {Schefer}, {Apostolakos},
  {{\'A}lvarez-Ayll{\'o}n}, {Dubath}  \& {K{\"u}mmel}}{{Bertin}
  et~al.}{2020}]{bertin20}
{Bertin} E.,  {Schefer} M.,  {Apostolakos} N.,  {{\'A}lvarez-Ayll{\'o}n} A.,
  {Dubath} P.,   {K{\"u}mmel} M.,  2020, in {Pizzo} R.,  {Deul} E.~R.,  {Mol}
  J.~D.,  {de Plaa} J.,   {Verkouter} H.,  eds,  Astronomical Society of the
  Pacific Conference Series Vol. 527, Astronomical Data Analysis Software and
  Systems XXIX. p.~461

\bibitem[\protect\citeauthoryear{{Berzin}, {Secunda}, {Cen}, {Menegas}  \&
  {G{\"o}tberg}}{{Berzin} et~al.}{2021}]{berzin21}
{Berzin} E.,  {Secunda} A.,  {Cen} R.,  {Menegas} A.,   {G{\"o}tberg} Y.,
  2021, \mn@doi [\apj] {10.3847/1538-4357/ac0af6}, \href
  {https://ui.adsabs.harvard.edu/abs/2021ApJ...918....5B} {918, 5}

\bibitem[\protect\citeauthoryear{{Bhatawdekar}, {Conselice},
  {Margalef-Bentabol}  \& {Duncan}}{{Bhatawdekar} et~al.}{2019}]{bhatawdekar19}
{Bhatawdekar} R.,  {Conselice} C.~J.,  {Margalef-Bentabol} B.,   {Duncan} K.,
  2019, \mn@doi [\mnras] {10.1093/mnras/stz866}, \href
  {https://ui.adsabs.harvard.edu/abs/2019MNRAS.486.3805B} {486, 3805}

\bibitem[\protect\citeauthoryear{{Bouchet} et~al.,}{{Bouchet}
  et~al.}{2015}]{bouchet12}
{Bouchet} P.,  et~al., 2015, \mn@doi [\pasp] {10.1086/682254}, \href
  {https://ui.adsabs.harvard.edu/abs/2015PASP..127..612B} {127, 612}

\bibitem[\protect\citeauthoryear{{Bouwens} et~al.,}{{Bouwens}
  et~al.}{2011}]{bouwens11}
{Bouwens} R.~J.,  et~al., 2011, \mn@doi [\apj] {10.1088/0004-637X/737/2/90},
  \href {http://adsabs.harvard.edu/abs/2011ApJ...737...90B} {737, 90}

\bibitem[\protect\citeauthoryear{{Bouwens}, {Oesch}, {Illingworth}, {Ellis}  \&
  {Stefanon}}{{Bouwens} et~al.}{2017}]{bouwens17}
{Bouwens} R.~J.,  {Oesch} P.~A.,  {Illingworth} G.~D.,  {Ellis} R.~S.,
  {Stefanon} M.,  2017, \mn@doi [\apj] {10.3847/1538-4357/aa70a4}, \href
  {http://adsabs.harvard.edu/abs/2017ApJ...843..129B} {843, 129}

\bibitem[\protect\citeauthoryear{{Bouwens}, {Illingworth}, {Ellis}, {Oesch},
  {Paulino-Afonso}, {Ribeiro}  \& {Stefanon}}{{Bouwens}
  et~al.}{2022a}]{bouwens22a}
{Bouwens} R.~J.,  {Illingworth} G.,  {Ellis} R.~S.,  {Oesch} P.,
  {Paulino-Afonso} A.,  {Ribeiro} B.,   {Stefanon} M.,  2022a, \mn@doi [\apj]
  {10.3847/1538-4357/ac618c}, \href
  {https://ui.adsabs.harvard.edu/abs/2022ApJ...931...81B} {931, 81}

\bibitem[\protect\citeauthoryear{{Bouwens}, {Illingworth}, {Ellis}, {Oesch}  \&
  {Stefanon}}{{Bouwens} et~al.}{2022b}]{bouwens22b}
{Bouwens} R.~J.,  {Illingworth} G.,  {Ellis} R.~S.,  {Oesch} P.,   {Stefanon}
  M.,  2022b, \mn@doi [\apj] {10.3847/1538-4357/ac86d1}, \href
  {https://ui.adsabs.harvard.edu/abs/2022ApJ...940...55B} {940, 55}

\bibitem[\protect\citeauthoryear{{Boylan-Kolchin}}{{Boylan-Kolchin}}{2022}]{boylan-kolchin22}
{Boylan-Kolchin} M.,  2022, arXiv e-prints, \href
  {https://ui.adsabs.harvard.edu/abs/2022arXiv220801611B} {p. arXiv:2208.01611}

\bibitem[\protect\citeauthoryear{{Brammer}, {van Dokkum}  \& {Coppi}}{{Brammer}
  et~al.}{2008}]{brammer08}
{Brammer} G.~B.,  {van Dokkum} P.~G.,   {Coppi} P.,  2008, \mn@doi [\apj]
  {10.1086/591786}, \href
  {https://ui.adsabs.harvard.edu/abs/2008ApJ...686.1503B} {686, 1503}

\bibitem[\protect\citeauthoryear{{Bruzual} \& {Charlot}}{{Bruzual} \&
  {Charlot}}{2003}]{bc03}
{Bruzual} G.,  {Charlot} S.,  2003, \mn@doi [\mnras]
  {10.1046/j.1365-8711.2003.06897.x}, \href
  {http://adsabs.harvard.edu/abs/2003MNRAS.344.1000B} {344, 1000}

\bibitem[\protect\citeauthoryear{{Calzetti}}{{Calzetti}}{2013}]{calzetti13}
{Calzetti} D.,  2013, in {Falc{\'o}n-Barroso} J.,  {Knapen} J.~H.,  eds, ,
  Secular Evolution of Galaxies.
p.~419

\bibitem[\protect\citeauthoryear{{Caminha}, {Suyu}, {Mercurio}, {Brammer},
  {Bergamini}, {Acebron}  \& {Vanzella}}{{Caminha} et~al.}{2022}]{caminha22}
{Caminha} G.~B.,  {Suyu} S.~H.,  {Mercurio} A.,  {Brammer} G.,  {Bergamini} P.,
   {Acebron} A.,   {Vanzella} E.,  2022, \mn@doi [\aap]
  {10.1051/0004-6361/202244517}, \href
  {https://ui.adsabs.harvard.edu/abs/2022A&A...666L...9C} {666, L9}

\bibitem[\protect\citeauthoryear{{Capak} et~al.,}{{Capak}
  et~al.}{2015}]{capak15}
{Capak} P.~L.,  et~al., 2015, \mn@doi [\nat] {10.1038/nature14500}, \href
  {https://ui.adsabs.harvard.edu/abs/2015Natur.522..455C} {522, 455}

\bibitem[\protect\citeauthoryear{{Carnall} et~al.,}{{Carnall}
  et~al.}{2023}]{carnall23}
{Carnall} A.~C.,  et~al., 2023, \mn@doi [\mnras] {10.1093/mnrasl/slac136},
  \href {https://ui.adsabs.harvard.edu/abs/2023MNRAS.518L..45C} {518, L45}

\bibitem[\protect\citeauthoryear{{Cassata} et~al.,}{{Cassata}
  et~al.}{2013}]{cassata13}
{Cassata} P.,  et~al., 2013, \mn@doi [\aap] {10.1051/0004-6361/201220969},
  \href {https://ui.adsabs.harvard.edu/abs/2013A&A...556A..68C} {556, A68}

\bibitem[\protect\citeauthoryear{{Castellano} et~al.,}{{Castellano}
  et~al.}{2022}]{castellano22}
{Castellano} M.,  et~al., 2022, \mn@doi [\apjl] {10.3847/2041-8213/ac94d0},
  \href {https://ui.adsabs.harvard.edu/abs/2022ApJ...938L..15C} {938, L15}

\bibitem[\protect\citeauthoryear{{Chabrier}}{{Chabrier}}{2003}]{chabrier03}
{Chabrier} G.,  2003, \mn@doi [\pasp] {10.1086/376392}, \href
  {https://ui.adsabs.harvard.edu/abs/2003PASP..115..763C} {115, 763}

\bibitem[\protect\citeauthoryear{{Chevallard} \& {Charlot}}{{Chevallard} \&
  {Charlot}}{2016}]{chevallard16}
{Chevallard} J.,  {Charlot} S.,  2016, \mn@doi [\mnras]
  {10.1093/mnras/stw1756}, \href
  {https://ui.adsabs.harvard.edu/abs/2016MNRAS.462.1415C} {462, 1415}

\bibitem[\protect\citeauthoryear{{Chisholm} et~al.,}{{Chisholm}
  et~al.}{2022}]{chisholm22}
{Chisholm} J.,  et~al., 2022, \mn@doi [\mnras] {10.1093/mnras/stac2874}, \href
  {https://ui.adsabs.harvard.edu/abs/2022MNRAS.517.5104C} {517, 5104}

\bibitem[\protect\citeauthoryear{{Coe} et~al.,}{{Coe} et~al.}{2019}]{coe19}
{Coe} D.,  et~al., 2019, \mn@doi [\apj] {10.3847/1538-4357/ab412b}, \href
  {https://ui.adsabs.harvard.edu/abs/2019ApJ...884...85C} {884, 85}

\bibitem[\protect\citeauthoryear{{Cullen} et~al.,}{{Cullen}
  et~al.}{2022}]{cullen22}
{Cullen} F.,  et~al., 2022, arXiv e-prints, \href
  {https://ui.adsabs.harvard.edu/abs/2022arXiv220804914C} {p. arXiv:2208.04914}

\bibitem[\protect\citeauthoryear{{Daddi} et~al.,}{{Daddi}
  et~al.}{2007}]{daddi07}
{Daddi} E.,  et~al., 2007, \mn@doi [\apj] {10.1086/521818}, \href
  {https://ui.adsabs.harvard.edu/abs/2007ApJ...670..156D} {670, 156}

\bibitem[\protect\citeauthoryear{{Dom{\'\i}nguez}, {Siana}, {Brooks},
  {Christensen}, {Bruzual}, {Stark}  \& {Alavi}}{{Dom{\'\i}nguez}
  et~al.}{2015}]{dominguez15}
{Dom{\'\i}nguez} A.,  {Siana} B.,  {Brooks} A.~M.,  {Christensen} C.~R.,
  {Bruzual} G.,  {Stark} D.~P.,   {Alavi} A.,  2015, \mn@doi [\mnras]
  {10.1093/mnras/stv1001}, \href
  {https://ui.adsabs.harvard.edu/abs/2015MNRAS.451..839D} {451, 839}

\bibitem[\protect\citeauthoryear{{Donnan} et~al.,}{{Donnan}
  et~al.}{2023}]{donnan23}
{Donnan} C.~T.,  et~al., 2023, \mn@doi [\mnras] {10.1093/mnras/stac3472}, \href
  {https://ui.adsabs.harvard.edu/abs/2023MNRAS.518.6011D} {518, 6011}

\bibitem[\protect\citeauthoryear{{Doyon} et~al.,}{{Doyon}
  et~al.}{2012}]{doyon12}
{Doyon} R.,  et~al., 2012, in {Clampin} M.~C.,  {Fazio} G.~G.,  {MacEwen}
  H.~A.,   {Oschmann} Jacobus~M. J.,  eds,  Society of Photo-Optical
  Instrumentation Engineers (SPIE) Conference Series Vol. 8442, Space
  Telescopes and Instrumentation 2012: Optical, Infrared, and Millimeter Wave.
  p. 84422R, \mn@doi{10.1117/12.926578}

\bibitem[\protect\citeauthoryear{{Elbaz} et~al.,}{{Elbaz}
  et~al.}{2007}]{elbaz07}
{Elbaz} D.,  et~al., 2007, \mn@doi [\aap] {10.1051/0004-6361:20077525}, \href
  {https://ui.adsabs.harvard.edu/abs/2007A&A...468...33E} {468, 33}

\bibitem[\protect\citeauthoryear{{Emami}, {Siana}, {Weisz}, {Johnson}, {Ma}  \&
  {El-Badry}}{{Emami} et~al.}{2019}]{emami19}
{Emami} N.,  {Siana} B.,  {Weisz} D.~R.,  {Johnson} B.~D.,  {Ma} X.,
  {El-Badry} K.,  2019, \mn@doi [\apj] {10.3847/1538-4357/ab211a}, \href
  {https://ui.adsabs.harvard.edu/abs/2019ApJ...881...71E} {881, 71}

\bibitem[\protect\citeauthoryear{{Endsley}, {Stark}, {Chevallard}  \&
  {Charlot}}{{Endsley} et~al.}{2021}]{endsley21}
{Endsley} R.,  {Stark} D.~P.,  {Chevallard} J.,   {Charlot} S.,  2021, \mn@doi
  [\mnras] {10.1093/mnras/staa3370}, \href
  {https://ui.adsabs.harvard.edu/abs/2021MNRAS.500.5229E} {500, 5229}

\bibitem[\protect\citeauthoryear{{Ferland} et~al.,}{{Ferland}
  et~al.}{2013}]{ferland13}
{Ferland} G.~J.,  et~al., 2013, \rmxaa, \href
  {https://ui.adsabs.harvard.edu/abs/2013RMxAA..49..137F} {49, 137}

\bibitem[\protect\citeauthoryear{{Ferrara}, {Pallottini}  \& {Dayal}}{{Ferrara}
  et~al.}{2022}]{ferrara22}
{Ferrara} A.,  {Pallottini} A.,   {Dayal} P.,  2022, arXiv e-prints, \href
  {https://ui.adsabs.harvard.edu/abs/2022arXiv220800720F} {p. arXiv:2208.00720}

\bibitem[\protect\citeauthoryear{{Finkelstein} et~al.,}{{Finkelstein}
  et~al.}{2022a}]{finkelstein23}
{Finkelstein} S.~L.,  et~al., 2022a, arXiv e-prints, \href
  {https://ui.adsabs.harvard.edu/abs/2022arXiv221105792F} {p. arXiv:2211.05792}

\bibitem[\protect\citeauthoryear{{Finkelstein} et~al.,}{{Finkelstein}
  et~al.}{2022b}]{finkelstein22}
{Finkelstein} S.~L.,  et~al., 2022b, \mn@doi [\apjl]
  {10.3847/2041-8213/ac966e}, \href
  {https://ui.adsabs.harvard.edu/abs/2022ApJ...940L..55F} {940, L55}

\bibitem[\protect\citeauthoryear{{Foreman-Mackey}, {Hogg}, {Lang}  \&
  {Goodman}}{{Foreman-Mackey} et~al.}{2013}]{foreman-mackey13}
{Foreman-Mackey} D.,  {Hogg} D.~W.,  {Lang} D.,   {Goodman} J.,  2013, \mn@doi
  [\pasp] {10.1086/670067}, \href
  {https://ui.adsabs.harvard.edu/abs/2013PASP..125..306F} {125, 306}

\bibitem[\protect\citeauthoryear{{F{\"o}rster Schreiber} \&
  {Wuyts}}{{F{\"o}rster Schreiber} \& {Wuyts}}{2020}]{foerster-schreiber20}
{F{\"o}rster Schreiber} N.~M.,  {Wuyts} S.,  2020, \mn@doi [\araa]
  {10.1146/annurev-astro-032620-021910}, \href
  {https://ui.adsabs.harvard.edu/abs/2020ARA&A..58..661F} {58, 661}

\bibitem[\protect\citeauthoryear{{Fudamoto}, {Inoue}  \& {Sugahara}}{{Fudamoto}
  et~al.}{2022}]{fudamoto22}
{Fudamoto} Y.,  {Inoue} A.~K.,   {Sugahara} Y.,  2022, \mn@doi [\apjl]
  {10.3847/2041-8213/ac982b}, \href
  {https://ui.adsabs.harvard.edu/abs/2022ApJ...938L..24F} {938, L24}

\bibitem[\protect\citeauthoryear{{Fujimoto} et~al.,}{{Fujimoto}
  et~al.}{2022}]{fujimoto22}
{Fujimoto} S.,  et~al., 2022, arXiv e-prints, \href
  {https://ui.adsabs.harvard.edu/abs/2022arXiv221103896F} {p. arXiv:2211.03896}

\bibitem[\protect\citeauthoryear{{Furtak}, {Atek}, {Lehnert}, {Chevallard}  \&
  {Charlot}}{{Furtak} et~al.}{2021}]{furtak21}
{Furtak} L.~J.,  {Atek} H.,  {Lehnert} M.~D.,  {Chevallard} J.,   {Charlot} S.,
   2021, \mn@doi [\mnras] {10.1093/mnras/staa3760}, \href
  {https://ui.adsabs.harvard.edu/abs/2021MNRAS.501.1568F} {501, 1568}

\bibitem[\protect\citeauthoryear{{Furtak} et~al.,}{{Furtak}
  et~al.}{2022}]{furtak22}
{Furtak} L.~J.,  et~al., 2022, \mn@doi [\mnras] {10.1093/mnras/stac2169}, \href
  {https://ui.adsabs.harvard.edu/abs/2022MNRAS.516.1373F} {516, 1373}

\bibitem[\protect\citeauthoryear{{Glazebrook} et~al.,}{{Glazebrook}
  et~al.}{2022}]{glazebrook22}
{Glazebrook} K.,  et~al., 2022, arXiv e-prints, \href
  {https://ui.adsabs.harvard.edu/abs/2022arXiv220803468G} {p. arXiv:2208.03468}

\bibitem[\protect\citeauthoryear{{Golubchik}, {Furtak}, {Meena}  \&
  {Zitrin}}{{Golubchik} et~al.}{2022}]{golubchik22}
{Golubchik} M.,  {Furtak} L.~J.,  {Meena} A.~K.,   {Zitrin} A.,  2022, \mn@doi
  [\apj] {10.3847/1538-4357/ac8ff1}, \href
  {https://ui.adsabs.harvard.edu/abs/2022ApJ...938...14G} {938, 14}

\bibitem[\protect\citeauthoryear{{Grazian} et~al.,}{{Grazian}
  et~al.}{2015}]{grazian15}
{Grazian} A.,  et~al., 2015, \mn@doi [\aap] {10.1051/0004-6361/201424750},
  \href {https://ui.adsabs.harvard.edu/abs/2015A&A...575A..96G} {575, A96}

\bibitem[\protect\citeauthoryear{{Gutkin}, {Charlot}  \& {Bruzual}}{{Gutkin}
  et~al.}{2016}]{gutkin16}
{Gutkin} J.,  {Charlot} S.,   {Bruzual} G.,  2016, \mn@doi [\mnras]
  {10.1093/mnras/stw1716}, \href
  {https://ui.adsabs.harvard.edu/abs/2016MNRAS.462.1757G} {462, 1757}

\bibitem[\protect\citeauthoryear{{Hao}, {Kennicutt}, {Johnson}, {Calzetti},
  {Dale}  \& {Moustakas}}{{Hao} et~al.}{2011}]{hao11}
{Hao} C.-N.,  {Kennicutt} R.~C.,  {Johnson} B.~D.,  {Calzetti} D.,  {Dale}
  D.~A.,   {Moustakas} J.,  2011, \mn@doi [\apj] {10.1088/0004-637X/741/2/124},
  \href {https://ui.adsabs.harvard.edu/abs/2011ApJ...741..124H} {741, 124}

\bibitem[\protect\citeauthoryear{{Harikane} et~al.,}{{Harikane}
  et~al.}{2022}]{harikane22}
{Harikane} Y.,  et~al., 2022, arXiv e-prints, \href
  {https://ui.adsabs.harvard.edu/abs/2022arXiv220801612H} {p. arXiv:2208.01612}

\bibitem[\protect\citeauthoryear{Hunter}{Hunter}{2007}]{hunter07}
Hunter J.~D.,  2007, \mn@doi [Computing in Science \& Engineering]
  {10.1109/MCSE.2007.55}, 9, 90

\bibitem[\protect\citeauthoryear{{Inoue}, {Shimizu}, {Iwata}  \&
  {Tanaka}}{{Inoue} et~al.}{2014}]{inoue14}
{Inoue} A.~K.,  {Shimizu} I.,  {Iwata} I.,   {Tanaka} M.,  2014, \mn@doi
  [\mnras] {10.1093/mnras/stu936}, \href
  {https://ui.adsabs.harvard.edu/abs/2014MNRAS.442.1805I} {442, 1805}

\bibitem[\protect\citeauthoryear{{Jakobsen} et~al.,}{{Jakobsen}
  et~al.}{2022}]{jakobsen22}
{Jakobsen} P.,  et~al., 2022, \mn@doi [\aap] {10.1051/0004-6361/202142663},
  \href {https://ui.adsabs.harvard.edu/abs/2022A&A...661A..80J} {661, A80}

\bibitem[\protect\citeauthoryear{{Jullo} \& {Kneib}}{{Jullo} \&
  {Kneib}}{2009}]{jullo09}
{Jullo} E.,  {Kneib} J.-P.,  2009, \mn@doi [\mnras]
  {10.1111/j.1365-2966.2009.14654.x}, \href
  {http://cdsads.u-strasbg.fr/abs/2009MNRAS.395.1319J} {395, 1319}

\bibitem[\protect\citeauthoryear{{Jullo}, {Kneib}, {Limousin},
  {El{\'{\i}}asd{\'o}ttir}, {Marshall}  \& {Verdugo}}{{Jullo}
  et~al.}{2007}]{jullo07}
{Jullo} E.,  {Kneib} J.-P.,  {Limousin} M.,  {El{\'{\i}}asd{\'o}ttir} {\'A}.,
  {Marshall} P.~J.,   {Verdugo} T.,  2007, \mn@doi [New Journal of Physics]
  {10.1088/1367-2630/9/12/447}, \href
  {http://adsabs.harvard.edu/abs/2007NJPh....9..447J} {9, 447}

\bibitem[\protect\citeauthoryear{{Katz}, {Kimm}, {Ellis}, {Devriendt}  \&
  {Slyz}}{{Katz} et~al.}{2022}]{katz22a}
{Katz} H.,  {Kimm} T.,  {Ellis} R.~S.,  {Devriendt} J.,   {Slyz} A.,  2022,
  arXiv e-prints, \href {https://ui.adsabs.harvard.edu/abs/2022arXiv220704751K}
  {p. arXiv:2207.04751}

\bibitem[\protect\citeauthoryear{{Kennicutt}}{{Kennicutt}}{1998}]{kennicutt98}
{Kennicutt} Robert~C. J.,  1998, \mn@doi [\apj] {10.1086/305588}, \href
  {https://ui.adsabs.harvard.edu/abs/1998ApJ...498..541K} {498, 541}

\bibitem[\protect\citeauthoryear{{Kennicutt} \& {Evans}}{{Kennicutt} \&
  {Evans}}{2012}]{kennicutt12}
{Kennicutt} R.~C.,  {Evans} N.~J.,  2012, \mn@doi [\araa]
  {10.1146/annurev-astro-081811-125610}, \href
  {https://ui.adsabs.harvard.edu/abs/2012ARA&A..50..531K} {50, 531}

\bibitem[\protect\citeauthoryear{{Kikuchihara} et~al.,}{{Kikuchihara}
  et~al.}{2020}]{kikuchihara20}
{Kikuchihara} S.,  et~al., 2020, \mn@doi [\apj] {10.3847/1538-4357/ab7dbe},
  \href {https://ui.adsabs.harvard.edu/abs/2020ApJ...893...60K} {893, 60}

\bibitem[\protect\citeauthoryear{{Kneib}, {Ellis}, {Smail}, {Couch}  \&
  {Sharples}}{{Kneib} et~al.}{1996}]{kneib96}
{Kneib} J.-P.,  {Ellis} R.~S.,  {Smail} I.,  {Couch} W.~J.,   {Sharples} R.~M.,
   1996, \mn@doi [\apj] {10.1086/177995}, \href
  {https://ui.adsabs.harvard.edu/abs/1996ApJ...471..643K} {471, 643}

\bibitem[\protect\citeauthoryear{{K{\"u}mmel}, {Bertin}, {Schefer},
  {Apostolakos}, {{\'A}lvarez-Ayll{\'o}n}  \& {Dubath}}{{K{\"u}mmel}
  et~al.}{2020}]{kuemmel20}
{K{\"u}mmel} M.,  {Bertin} E.,  {Schefer} M.,  {Apostolakos} N.,
  {{\'A}lvarez-Ayll{\'o}n} A.,   {Dubath} P.,  2020, in {Pizzo} R.,  {Deul}
  E.~R.,  {Mol} J.~D.,  {de Plaa} J.,   {Verkouter} H.,  eds,  Astronomical
  Society of the Pacific Conference Series Vol. 527, Astronomical Data Analysis
  Software and Systems XXIX. p.~29

\bibitem[\protect\citeauthoryear{{Labbe} et~al.,}{{Labbe}
  et~al.}{2022}]{labbe22}
{Labbe} I.,  et~al., 2022, arXiv e-prints, \href
  {https://ui.adsabs.harvard.edu/abs/2022arXiv220712446L} {p. arXiv:2207.12446}

\bibitem[\protect\citeauthoryear{{Laporte}, {Meyer}, {Ellis}, {Robertson},
  {Chisholm}  \& {Roberts-Borsani}}{{Laporte} et~al.}{2021}]{laporte21a}
{Laporte} N.,  {Meyer} R.~A.,  {Ellis} R.~S.,  {Robertson} B.~E.,  {Chisholm}
  J.,   {Roberts-Borsani} G.~W.,  2021, \mn@doi [\mnras]
  {10.1093/mnras/stab1239}, \href
  {https://ui.adsabs.harvard.edu/abs/2021MNRAS.505.3336L} {505, 3336}

\bibitem[\protect\citeauthoryear{{Laporte}, {Zitrin}, {Dole},
  {Roberts-Borsani}, {Furtak}  \& {Witten}}{{Laporte} et~al.}{2022}]{laporte22}
{Laporte} N.,  {Zitrin} A.,  {Dole} H.,  {Roberts-Borsani} G.,  {Furtak} L.~J.,
    {Witten} C.,  2022, \mn@doi [\aap] {10.1051/0004-6361/202244719}, \href
  {https://ui.adsabs.harvard.edu/abs/2022A&A...667L...3L} {667, L3}

\bibitem[\protect\citeauthoryear{{Lehnert} \& {Heckman}}{{Lehnert} \&
  {Heckman}}{1996}]{lehnert96}
{Lehnert} M.~D.,  {Heckman} T.~M.,  1996, \mn@doi [\apj] {10.1086/178086},
  \href {https://ui.adsabs.harvard.edu/abs/1996ApJ...472..546L} {472, 546}

\bibitem[\protect\citeauthoryear{{Leitherer} et~al.,}{{Leitherer}
  et~al.}{1999}]{leitherer99}
{Leitherer} C.,  et~al., 1999, \mn@doi [\apjs] {10.1086/313233}, \href
  {https://ui.adsabs.harvard.edu/abs/1999ApJS..123....3L} {123, 3}

\bibitem[\protect\citeauthoryear{{Ling} et~al.,}{{Ling} et~al.}{2022}]{ling22}
{Ling} C.-T.,  et~al., 2022, \mn@doi [\mnras] {10.1093/mnras/stac2716}, \href
  {https://ui.adsabs.harvard.edu/abs/2022MNRAS.517..853L} {517, 853}

\bibitem[\protect\citeauthoryear{{Mahler} et~al.,}{{Mahler}
  et~al.}{2022}]{mahler22}
{Mahler} G.,  et~al., 2022, arXiv e-prints, \href
  {https://ui.adsabs.harvard.edu/abs/2022arXiv220707101M} {p. arXiv:2207.07101}

\bibitem[\protect\citeauthoryear{{Mason}, {Trenti}  \& {Treu}}{{Mason}
  et~al.}{2015}]{mason15}
{Mason} C.~A.,  {Trenti} M.,   {Treu} T.,  2015, \mn@doi [\apj]
  {10.1088/0004-637X/813/1/21}, \href
  {https://ui.adsabs.harvard.edu/abs/2015ApJ...813...21M} {813, 21}

\bibitem[\protect\citeauthoryear{{Mason}, {Trenti}  \& {Treu}}{{Mason}
  et~al.}{2022}]{mason22}
{Mason} C.~A.,  {Trenti} M.,   {Treu} T.,  2022, arXiv e-prints, \href
  {https://ui.adsabs.harvard.edu/abs/2022arXiv220714808M} {p. arXiv:2207.14808}

\bibitem[\protect\citeauthoryear{{Meurer}, {Heckman}  \& {Calzetti}}{{Meurer}
  et~al.}{1999}]{meurer99}
{Meurer} G.~R.,  {Heckman} T.~M.,   {Calzetti} D.,  1999, \mn@doi [\apj]
  {10.1086/307523}, \href
  {https://ui.adsabs.harvard.edu/abs/1999ApJ...521...64M} {521, 64}

\bibitem[\protect\citeauthoryear{{Naidu} et~al.,}{{Naidu}
  et~al.}{2022a}]{naidu22c}
{Naidu} R.~P.,  et~al., 2022a, arXiv e-prints, \href
  {https://ui.adsabs.harvard.edu/abs/2022arXiv220802794N} {p. arXiv:2208.02794}

\bibitem[\protect\citeauthoryear{{Naidu} et~al.,}{{Naidu}
  et~al.}{2022b}]{naidu22b}
{Naidu} R.~P.,  et~al., 2022b, \mn@doi [\apjl] {10.3847/2041-8213/ac9b22},
  \href {https://ui.adsabs.harvard.edu/abs/2022ApJ...940L..14N} {940, L14}

\bibitem[\protect\citeauthoryear{{Nanayakkara} et~al.,}{{Nanayakkara}
  et~al.}{2022}]{nanayakkara22}
{Nanayakkara} T.,  et~al., 2022, arXiv e-prints, \href
  {https://ui.adsabs.harvard.edu/abs/2022arXiv220713860N} {p. arXiv:2207.13860}

\bibitem[\protect\citeauthoryear{{Nelson} et~al.,}{{Nelson}
  et~al.}{2022}]{nelson22}
{Nelson} E.~J.,  et~al., 2022, arXiv e-prints, \href
  {https://ui.adsabs.harvard.edu/abs/2022arXiv220801630N} {p. arXiv:2208.01630}

\bibitem[\protect\citeauthoryear{{Noeske} et~al.,}{{Noeske}
  et~al.}{2007}]{noeske07}
{Noeske} K.~G.,  et~al., 2007, \mn@doi [\apjl] {10.1086/517926}, \href
  {https://ui.adsabs.harvard.edu/abs/2007ApJ...660L..43N} {660, L43}

\bibitem[\protect\citeauthoryear{{Nonino} et~al.,}{{Nonino}
  et~al.}{2022}]{nonino22}
{Nonino} M.,  et~al., 2022, arXiv e-prints, \href
  {https://ui.adsabs.harvard.edu/abs/2022arXiv220714802N} {p. arXiv:2207.14802}

\bibitem[\protect\citeauthoryear{{Oke} \& {Gunn}}{{Oke} \&
  {Gunn}}{1983}]{oke83}
{Oke} J.~B.,  {Gunn} J.~E.,  1983, \mn@doi [\apj] {10.1086/160817}, \href
  {http://adsabs.harvard.edu/abs/1983ApJ...266..713O} {266, 713}

\bibitem[\protect\citeauthoryear{{Pascale} et~al.,}{{Pascale}
  et~al.}{2022}]{pascale22}
{Pascale} M.,  et~al., 2022, \mn@doi [\apjl] {10.3847/2041-8213/ac9316}, \href
  {https://ui.adsabs.harvard.edu/abs/2022ApJ...938L...6P} {938, L6}

\bibitem[\protect\citeauthoryear{{Pei}}{{Pei}}{1992}]{pei92}
{Pei} Y.~C.,  1992, \mn@doi [\apj] {10.1086/171637}, \href
  {https://ui.adsabs.harvard.edu/abs/1992ApJ...395..130P} {395, 130}

\bibitem[\protect\citeauthoryear{{Plat}, {Charlot}, {Bruzual}, {Feltre},
  {Vidal-Garc{\'\i}a}, {Morisset}, {Chevallard}  \& {Todt}}{{Plat}
  et~al.}{2019}]{plat19}
{Plat} A.,  {Charlot} S.,  {Bruzual} G.,  {Feltre} A.,  {Vidal-Garc{\'\i}a} A.,
   {Morisset} C.,  {Chevallard} J.,   {Todt} H.,  2019, \mn@doi [\mnras]
  {10.1093/mnras/stz2616}, \href
  {https://ui.adsabs.harvard.edu/abs/2019MNRAS.490..978P} {490, 978}

\bibitem[\protect\citeauthoryear{{Pontoppidan} et~al.,}{{Pontoppidan}
  et~al.}{2022}]{pontoppidan22}
{Pontoppidan} K.~M.,  et~al., 2022, \mn@doi [\apjl] {10.3847/2041-8213/ac8a4e},
  \href {https://ui.adsabs.harvard.edu/abs/2022ApJ...936L..14P} {936, L14}

\bibitem[\protect\citeauthoryear{{Price-Whelan} et~al.,}{{Price-Whelan}
  et~al.}{2018}]{astropy18}
{Price-Whelan} A.~M.,  et~al., 2018, \mn@doi [\aj] {10.3847/1538-3881/aabc4f},
  \href {https://ui.adsabs.harvard.edu/#abs/2018AJ....156..123T} {156, 123}

\bibitem[\protect\citeauthoryear{{Reddy} et~al.,}{{Reddy}
  et~al.}{2015}]{reddy15}
{Reddy} N.~A.,  et~al., 2015, \mn@doi [\apj] {10.1088/0004-637X/806/2/259},
  \href {https://ui.adsabs.harvard.edu/abs/2015ApJ...806..259R} {806, 259}

\bibitem[\protect\citeauthoryear{{Reddy} et~al.,}{{Reddy}
  et~al.}{2018}]{reddy18a}
{Reddy} N.~A.,  et~al., 2018, \mn@doi [\apj] {10.3847/1538-4357/aaa3e7}, \href
  {https://ui.adsabs.harvard.edu/abs/2018ApJ...853...56R} {853, 56}

\bibitem[\protect\citeauthoryear{{Riaz}, {Hartwig}  \& {Latif}}{{Riaz}
  et~al.}{2022}]{riaz22}
{Riaz} S.,  {Hartwig} T.,   {Latif} M.~A.,  2022, \mn@doi [\apjl]
  {10.3847/2041-8213/ac8ea6}, \href
  {https://ui.adsabs.harvard.edu/abs/2022ApJ...937L...6R} {937, L6}

\bibitem[\protect\citeauthoryear{{Rieke}, {Kelly}  \& {Horner}}{{Rieke}
  et~al.}{2005}]{rieke05}
{Rieke} M.~J.,  {Kelly} D.,   {Horner} S.,  2005, {Overview of James Webb Space
  Telescope and NIRCam's Role}.
SPIE, pp~1--8, \mn@doi{10.1117/12.615554}

\bibitem[\protect\citeauthoryear{{Rieke} et~al.,}{{Rieke}
  et~al.}{2015}]{rieke15}
{Rieke} G.~H.,  et~al., 2015, \mn@doi [\pasp] {10.1086/682252}, \href
  {https://ui.adsabs.harvard.edu/abs/2015PASP..127..584R} {127, 584}

\bibitem[\protect\citeauthoryear{{Roberts-Borsani} et~al.,}{{Roberts-Borsani}
  et~al.}{2022}]{roberts-borsani22}
{Roberts-Borsani} G.,  et~al., 2022, arXiv e-prints, \href
  {https://ui.adsabs.harvard.edu/abs/2022arXiv221015639R} {p. arXiv:2210.15639}

\bibitem[\protect\citeauthoryear{{Rodighiero}, {Bisigello}, {Iani}, {Marasco},
  {Grazian}, {Sinigaglia}, {Cassata}  \& {Gruppioni}}{{Rodighiero}
  et~al.}{2023}]{rodighiero23}
{Rodighiero} G.,  {Bisigello} L.,  {Iani} E.,  {Marasco} A.,  {Grazian} A.,
  {Sinigaglia} F.,  {Cassata} P.,   {Gruppioni} C.,  2023, \mn@doi [\mnras]
  {10.1093/mnrasl/slac115}, \href
  {https://ui.adsabs.harvard.edu/abs/2023MNRAS.518L..19R} {518, L19}

\bibitem[\protect\citeauthoryear{{Sanati}, {Jeanquartier}, {Revaz}  \&
  {Jablonka}}{{Sanati} et~al.}{2022}]{sanati22}
{Sanati} M.,  {Jeanquartier} F.,  {Revaz} Y.,   {Jablonka} P.,  2022, arXiv
  e-prints, \href {https://ui.adsabs.harvard.edu/abs/2022arXiv220611351S} {p.
  arXiv:2206.11351}

\bibitem[\protect\citeauthoryear{{Schaerer}, {Marques-Chaves}, {Barrufet},
  {Oesch}, {Izotov}, {Naidu}, {Guseva}  \& {Brammer}}{{Schaerer}
  et~al.}{2022}]{schaerer22b}
{Schaerer} D.,  {Marques-Chaves} R.,  {Barrufet} L.,  {Oesch} P.,  {Izotov}
  Y.~I.,  {Naidu} R.,  {Guseva} N.~G.,   {Brammer} G.,  2022, \mn@doi [\aap]
  {10.1051/0004-6361/202244556}, \href
  {https://ui.adsabs.harvard.edu/abs/2022A&A...665L...4S} {665, L4}

\bibitem[\protect\citeauthoryear{{Sheth}, {Mo}  \& {Tormen}}{{Sheth}
  et~al.}{2001}]{sheth01}
{Sheth} R.~K.,  {Mo} H.~J.,   {Tormen} G.,  2001, \mn@doi [\mnras]
  {10.1046/j.1365-8711.2001.04006.x}, \href
  {https://ui.adsabs.harvard.edu/abs/2001MNRAS.323....1S} {323, 1}

\bibitem[\protect\citeauthoryear{{Shivaei} et~al.,}{{Shivaei}
  et~al.}{2020}]{shivaei20}
{Shivaei} I.,  et~al., 2020, \mn@doi [\apj] {10.3847/1538-4357/aba35e}, \href
  {https://ui.adsabs.harvard.edu/abs/2020ApJ...899..117S} {899, 117}

\bibitem[\protect\citeauthoryear{{Shuntov} et~al.,}{{Shuntov}
  et~al.}{2022}]{shuntov22}
{Shuntov} M.,  et~al., 2022, \mn@doi [\aap] {10.1051/0004-6361/202243136},
  \href {https://ui.adsabs.harvard.edu/abs/2022A&A...664A..61S} {664, A61}

\bibitem[\protect\citeauthoryear{{Song} et~al.,}{{Song} et~al.}{2016}]{song16}
{Song} M.,  et~al., 2016, \mn@doi [\apj] {10.3847/0004-637X/825/1/5}, \href
  {https://ui.adsabs.harvard.edu/abs/2016ApJ...825....5S} {825, 5}

\bibitem[\protect\citeauthoryear{{Speagle}, {Steinhardt}, {Capak}  \&
  {Silverman}}{{Speagle} et~al.}{2014}]{speagle14}
{Speagle} J.~S.,  {Steinhardt} C.~L.,  {Capak} P.~L.,   {Silverman} J.~D.,
  2014, \mn@doi [\apjs] {10.1088/0067-0049/214/2/15}, \href
  {https://ui.adsabs.harvard.edu/abs/2014ApJS..214...15S} {214, 15}

\bibitem[\protect\citeauthoryear{{Stark}, {Ellis}, {Bunker}, {Bundy},
  {Targett}, {Benson}  \& {Lacy}}{{Stark} et~al.}{2009}]{stark09}
{Stark} D.~P.,  {Ellis} R.~S.,  {Bunker} A.,  {Bundy} K.,  {Targett} T.,
  {Benson} A.,   {Lacy} M.,  2009, \mn@doi [\apj]
  {10.1088/0004-637X/697/2/1493}, \href
  {https://ui.adsabs.harvard.edu/abs/2009ApJ...697.1493S} {697, 1493}

\bibitem[\protect\citeauthoryear{{Stefanon}, {Bouwens}, {Labb{\'e}},
  {Illingworth}, {Gonzalez}  \& {Oesch}}{{Stefanon} et~al.}{2021}]{stefanon21}
{Stefanon} M.,  {Bouwens} R.~J.,  {Labb{\'e}} I.,  {Illingworth} G.~D.,
  {Gonzalez} V.,   {Oesch} P.~A.,  2021, \mn@doi [\apj]
  {10.3847/1538-4357/ac1bb6}, \href
  {https://ui.adsabs.harvard.edu/abs/2021ApJ...922...29S} {922, 29}

\bibitem[\protect\citeauthoryear{{Tacchella}, {Bose}, {Conroy}, {Eisenstein}
  \& {Johnson}}{{Tacchella} et~al.}{2018}]{tacchella18}
{Tacchella} S.,  {Bose} S.,  {Conroy} C.,  {Eisenstein} D.~J.,   {Johnson}
  B.~D.,  2018, \mn@doi [\apj] {10.3847/1538-4357/aae8e0}, \href
  {https://ui.adsabs.harvard.edu/abs/2018ApJ...868...92T} {868, 92}

\bibitem[\protect\citeauthoryear{{Topping}, {Stark}, {Endsley}, {Plat},
  {Whitler}, {Chen}  \& {Charlot}}{{Topping} et~al.}{2022}]{topping22}
{Topping} M.~W.,  {Stark} D.~P.,  {Endsley} R.,  {Plat} A.,  {Whitler} L.,
  {Chen} Z.,   {Charlot} S.,  2022, \mn@doi [\apj] {10.3847/1538-4357/aca522},
  \href {https://ui.adsabs.harvard.edu/abs/2022ApJ...941..153T} {941, 153}

\bibitem[\protect\citeauthoryear{{Treu} et~al.,}{{Treu} et~al.}{2022}]{treu22}
{Treu} T.,  et~al., 2022, \mn@doi [\apj] {10.3847/1538-4357/ac8158}, \href
  {https://ui.adsabs.harvard.edu/abs/2022ApJ...935..110T} {935, 110}

\bibitem[\protect\citeauthoryear{{Verma}, {Lehnert}, {F{\"o}rster Schreiber},
  {Bremer}  \& {Douglas}}{{Verma} et~al.}{2007}]{verma07}
{Verma} A.,  {Lehnert} M.~D.,  {F{\"o}rster Schreiber} N.~M.,  {Bremer} M.~N.,
   {Douglas} L.,  2007, \mn@doi [\mnras] {10.1111/j.1365-2966.2007.11455.x},
  \href {https://ui.adsabs.harvard.edu/abs/2007MNRAS.377.1024V} {377, 1024}

\bibitem[\protect\citeauthoryear{{Virtanen} et~al.,}{{Virtanen}
  et~al.}{2020}]{virtanen20}
{Virtanen} P.,  et~al., 2020, \mn@doi [Nature Methods]
  {https://doi.org/10.1038/s41592-019-0686-2}, \href {https://rdcu.be/b08Wh}
  {17, 261}

\bibitem[\protect\citeauthoryear{{Weisz} et~al.,}{{Weisz}
  et~al.}{2012}]{weisz12}
{Weisz} D.~R.,  et~al., 2012, \mn@doi [\apj] {10.1088/0004-637X/744/1/44},
  \href {https://ui.adsabs.harvard.edu/abs/2012ApJ...744...44W} {744, 44}

\bibitem[\protect\citeauthoryear{{Whitaker} et~al.,}{{Whitaker}
  et~al.}{2014}]{whitaker14}
{Whitaker} K.~E.,  et~al., 2014, \mn@doi [\apj] {10.1088/0004-637X/795/2/104},
  \href {https://ui.adsabs.harvard.edu/abs/2014ApJ...795..104W} {795, 104}

\bibitem[\protect\citeauthoryear{{Whitler}, {Endsley}, {Stark}, {Topping},
  {Chen}  \& {Charlot}}{{Whitler} et~al.}{2023}]{whitler23}
{Whitler} L.,  {Endsley} R.,  {Stark} D.~P.,  {Topping} M.,  {Chen} Z.,
  {Charlot} S.,  2023, \mn@doi [\mnras] {10.1093/mnras/stac3535}, \href
  {https://ui.adsabs.harvard.edu/abs/2023MNRAS.519..157W} {519, 157}

\bibitem[\protect\citeauthoryear{{Williams} et~al.,}{{Williams}
  et~al.}{2022}]{williams22}
{Williams} H.,  et~al., 2022, arXiv e-prints, \href
  {https://ui.adsabs.harvard.edu/abs/2022arXiv221015699W} {p. arXiv:2210.15699}

\bibitem[\protect\citeauthoryear{{Yan}, {Ma}, {Ling}, {Cheng}, {Huang}  \&
  {Zitrin}}{{Yan} et~al.}{2022}]{yan22}
{Yan} H.,  {Ma} Z.,  {Ling} C.,  {Cheng} C.,  {Huang} J.-s.,   {Zitrin} A.,
  2022, arXiv e-prints, \href
  {https://ui.adsabs.harvard.edu/abs/2022arXiv220711558Y} {p. arXiv:2207.11558}

\bibitem[\protect\citeauthoryear{{Zackrisson}, {Inoue}  \&
  {Jensen}}{{Zackrisson} et~al.}{2013}]{zackrisson13}
{Zackrisson} E.,  {Inoue} A.~K.,   {Jensen} H.,  2013, \mn@doi [\apj]
  {10.1088/0004-637X/777/1/39}, \href
  {https://ui.adsabs.harvard.edu/abs/2013ApJ...777...39Z} {777, 39}

\bibitem[\protect\citeauthoryear{{Zackrisson} et~al.,}{{Zackrisson}
  et~al.}{2017}]{zackrisson17}
{Zackrisson} E.,  et~al., 2017, \mn@doi [\apj] {10.3847/1538-4357/836/1/78},
  \href {https://ui.adsabs.harvard.edu/abs/2017ApJ...836...78Z} {836, 78}

\bibitem[\protect\citeauthoryear{{Zavala} et~al.,}{{Zavala}
  et~al.}{2022}]{zavala22}
{Zavala} J.~A.,  et~al., 2022, arXiv e-prints, \href
  {https://ui.adsabs.harvard.edu/abs/2022arXiv220801816Z} {p. arXiv:2208.01816}

\bibitem[\protect\citeauthoryear{{Zitrin} et~al.,}{{Zitrin}
  et~al.}{2015}]{zitrin15a}
{Zitrin} A.,  et~al., 2015, \mn@doi [\apj] {10.1088/0004-637X/801/1/44}, \href
  {https://ui.adsabs.harvard.edu/abs/2015ApJ...801...44Z} {801, 44}

\bibitem[\protect\citeauthoryear{{van der Walt}, {Colbert}  \&
  {Varoquaux}}{{van der Walt} et~al.}{2011}]{vanderwalt11}
{van der Walt} S.,  {Colbert} S.~C.,   {Varoquaux} G.,  2011, Computing in
  Science Engineering, 13, 22

\makeatother
\end{thebibliography}

\appendix

\end{document}